 \documentclass[jgrga]{AGUTeX}
\usepackage{graphicx}
\usepackage{graphics}
\usepackage{url}
\usepackage{epstopdf}
\usepackage{color}
\usepackage{bm}
\usepackage{times}

\newcommand{\icarus}{Icarus}

\def\lsim{\mathrel{\hbox{\rlap{\hbox{\lower4pt\hbox{$\sim$}}}\hbox{$<$}}}}
\def\gsim{\mathrel{\hbox{\rlap{\hbox{\lower4pt\hbox{$\sim$}}}\hbox{$>$}}}}

\def\ft2pi{\frac{1}{\left(\sqrt{2\pi}\right)^{3}}}

\def\d3{\mbox{d}}

\authorrunninghead{TEODORO et al.}
\titlerunninghead{}
\authoraddr{
$^1$~BAER, NASA Ames Research Center, Moffett Field, CA 94035-1000, USA (luis.f.teodoro@nasa.gov), 
$^2$~Institute for Computational Cosmology, Department of Physics, Durham University, Science Laboratories, South Road, Durham DH1 3LE, UK,
$^3$~Planetary Systems Branch, Space Sciences and Astrobiology Division, MS 245-3, NASA Ames Research Center,  Moffett Field, CA 94035-1000, USA, 
$^4$~Planetary Science Institute, 1700 E. Fort Lowell, Suite 106, Tucson, AZ 85719, USA,
$^5$~Johns Hopkins University Applied Physics Laboratory, Laurel, MD 20723, USA.
}
\begin{document}

\setkeys{Gin}{draft=false}

\title{{
How well do we know the polar hydrogen distribution on the Moon?
}
}

 \authors{
 L.F.A. Teodoro, \altaffilmark{1} 
 V.R. Eke, \altaffilmark{2} 
 R.C. Elphic, \altaffilmark{3} 
 W.C. Feldman, \altaffilmark{4} and 
 D.J. Lawrence, \altaffilmark{5}}
\altaffiltext{1}{BAER, Planetary Systems Branch, Space Sciences and Astrobiology Division, MS 245-3, NASA Ames Research Center, Moffett Field, CA 94035-1000, USA}
\altaffiltext{2}{Institute for Computational Cosmology, Department of Physics, Durham University, Science Laboratories, South Road, Durham DH1 3LE, UK}
 \altaffiltext{3}{Planetary Systems Branch, Space Sciences and Astrobiology Division, MS 245-3, NASA Ames Research Center,  Moffett Field, CA 94035-1000, USA}
\altaffiltext{4}{Planetary Science Institute, 1700 E. Fort Lowell, Suite 106, Tucson, AZ 85719, USA}
\altaffiltext{5}{Johns Hopkins University Applied Physics Laboratory, Laurel, MD 20723, USA} 
\begin{article}

\abstract{
A detailed comparison is made of results from the Lunar Prospector
Neutron Spectrometer (LPNS) and the Lunar Exploration Neutron Detector
Collimated Sensors for EpiThermal Neutrons (LEND CSETN). Using the
autocorrelation function and power spectrum of the polar count rate
maps produced by these experiments, it is shown that the LEND CSETN
has a footprint that is at least as big as would be expected for an
omni-directional detector at an orbital altitude of $50$ km. The
collimated flux into the field of view of the collimator is
negligible. Arguments put forward asserting otherwise are considered
and found wanting for various reasons. The maps of lunar polar
hydrogen with the highest contrast, i.e. spatial resolution, are those
resulting from pixon image reconstructions of the LPNS data. These
typically provide weight percentages of water equivalent hydrogen that
are accurate to $30\%$ within the polar craters.
}

\vspace{0.4in}
\section{Introduction}\label{sec:intro}
The presence and distribution of hydrogen near the lunar surface is a
matter of considerable interest \citep{wat61,arn79}. This ancient
surface, like that of 
Mercury, contains a 
record of the history of the
inner solar system, and the likely association of hydrogen with water
molecules can provide insights into the delivery and retention of
volatile molecules over the past few billion years \citep{law13,paige13,neu13}.

Remote sensing of the epithermal neutron flux coming from the lunar
surface provides a measure of the hydrogen abundance in the top metre
or so of the lunar regolith \citep{lin61,metz90,feld91}. Cosmic rays
interacting with 
nuclei in the regolith create energetic, fast neutrons that
subsequently evolve and lose energy through inelastic and elastic
collisions with other nuclei. Some of these neutrons escape into space
before losing enough energy to be reabsorbed into another nucleus, and
this leakage flux contains information about the nuclear content of
the upper regolith. Hydrogen provides a very effective
moderator of intermediate energy, epithermal neutrons that
predominantly lose energy through elastic scattering. Consequently,
the presence of hydrogen in the top metre of regolith leads to a
relatively low flux of epithermal neutrons leaking from the surface.

Pioneering work in this subject was performed by those working with the
Lunar Prospector Neutron Spectrometer (LPNS), who mapped the lunar
neutron flux at fast, epithermal and thermal energies
\citep{feld98a,rick98,feld98}. Fast neutrons provide a map of the 
mean atomic mass \citep{gasnault001}, while thermal neutrons identify regions with
higher abundances of neutron-absorbing nuclei such as iron, titanium,
gadolinium and samarium. A deficit of epithermal neutrons is seen over
the mare regions, because the lower energy epithermal neutrons are
sensitive to the neutron absorbing nuclei \citep{law06}. While this is
not important 
for the polar regions, which have a feldspathic composition
characteristic of the lunar highlands, when making a global hydrogen
map \cite{feld00a} introduced a quantity $epi^*$ to correct for the
effects of these non-hydrogen absorbers at low latitudes. Nearer the
poles, the main aspect of composition driving the epithermal
neutron count rate is hydrogen and the LPNS results showed
reduced polar epithermal neutron count rates, implying
the presence of polar hydrogen.

With the $\sim 45$ km footprint size of the omni-directional LPNS
\citep{maur04} and the
inevitable stochastic noise present in the data, it was difficult to
determine if the dips in count rate were associated with the
relatively small permanently shaded regions (PSRs) that might be expected to
host water ice deposits. Consequently, pixon image reconstruction
techniques \citep{pp93,eke01} were employed to enhance the
information that could be extracted from the data. Using the method
introduced in \cite{rick07}, \cite{eke09} were the first to show
that the data favoured a scenario where the hydrogen was, on average,
concentrated into the PSRs. This analysis was improved using updated
maps of the PSRs by \cite{luis10}, whose maps were used in the
targeting of the Lunar Crater Observation and Sensing Satellite
(LCROSS) in its successful bid to find water ice in the Cabeus crater
\citep{cola10}. 

NASA's Lunar Precursor Robotic Program was intended to ``pave the way
for eventual permanent human presence on the Moon'' \citep{chin07}.
The first mission of this program was the Lunar Reconnaissance Orbiter
(LRO), which employs ``six individual instruments to produce accurate
maps and high-resolution images of future landing sites, to assess
potential lunar resouces, and to characterize the radiation
environment'' \citep{chin07}. One of these instruments is the Lunar
Exploration Neutron Detector (LEND), with a primary objective being to
``determine hydrogen content of the subsurface at the polar regions
with spatial resolution of $10$ km and with sensitivity to
concentration variations of 100 parts per million at the poles''
\citep{chin07}. 
Rather than taking omni-directional measurements and using software to
enhance the resulting images, as was done with the LPNS, the LEND
Collimated Sensors for 
EpiThermal Neutrons (CSETN) represent an attempt at a hardware
solution to the challenge of making sharper maps of the lunar
epithermal neutron count rate. This was to be achieved using a
two-layer collimator with an outer layer of polyethylene to moderate
the neutrons and an inner layer of boron to absorb them \citep{mit08}.

Prior to launch, there was a study anticipating how the
LEND CSETN might perform. \cite{law10} used Monte Carlo modelling
based on experience gained from work with the LPNS to infer that the
neutron count rate through the small field of view of the collimator
was going to be a rather low $0.18$ neutrons per second. However, this
disagreed with the estimate from \cite{mit08}, who found a value of
$0.9$ per second.

Analyses of orbital LEND CSETN data have resulted in discordant
inferences concerning the behaviour of the collimator. \cite{mit10b}
claimed that the LEND CSETN was receiving ``about $1.9$'' collimated neutrons
per second. On the basis of this interpretation, \cite{mit10b}
concluded that epithermal neutron suppressions, and by implication
enhanced hydrogen concentrations, were not spatially coincident with
permanently shaded regions. In response, \cite{law11b} contended that
the LEND CSETN count rate was dominated by an uncollimated high energy
epithermal neutron component. As a consequence, \cite{law11b}
concluded that the LEND CSETN data did not support the polar hydrogen
distributions inferred by \cite{mit10b}. A more
comprehensive likelihood analysis of the time series data was
performed by \cite{eke12}, who considered the three different
components contributing to the LEND CSETN count rate: the lunar
collimated component, the lunar uncollimated component, i.e. neutrons
from ouside the collimator field of view on the Moon that scatter off
spacecraft material into the detector, and neutrons generated by
cosmic rays striking spacecraft material itself.
Taking into account the three different
components contributing to the LEND CSETN count rate and how they
should vary with longitude, latitude and spacecraft altitude,
\cite{eke12} showed that the collimated count rate represented less
than about $10\%$ of the lunar-derived neutrons, allowing for
potential systematic uncertainties. The uncollimated lunar neutrons,
which provide a spatially varying background, dominated the count
rate from the Moon. However, more than half of the LEND CSETN
count rate is derived from cosmic rays striking the
spacecraft itself \citep{eke12}, so fewer than $5\%$ of the detected
neutrons were actually lunar and collimated. \cite{eke12} determined
that $1\%$ is the most likely fraction of the detected neutrons that
are lunar and collimated, meaning that the effective footprint of the
LEND CSETN will be set by the uncollimated lunar background component
and is likely to be at least $\sim 50$ km in size.
More recently, a number of 
papers have appeared \citep{lit12a,san12,mit12,lit12b,boy12}
that contain assertions to the effect that the LEND CSETN is producing
a map with $10$ km spatial resolution. 

Given the importance for the planning of future missions, it is
imperative that the capability of the LEND CSETN is clarified for
decision-makers outside the field of planetary neutron studies. The
purpose of this paper is to determine empirically the instrumental spatial
resolution and the background contamination in the data,
and hence the ability to map hydrogen near the lunar poles, of 
the LEND CSETN. This will be achieved using techniques that are new to
planetary neutron spectroscopy, but well established in other
scientific fields. 

In the next section, two statistical measures will be introduced to
characterise the performance of a detector given the output map it
produces. These will then be applied in Section~\ref{sec:app} to
the data sets from both the LPNS and LEND CSETN in order to compare
the relative performance of these two detectors. The various arguments
put forward by authors in support of statements about the proper
functioning of the LEND CSETN are investigated in detail in
Section~\ref{sec:other}. The results from this study are discussed in
Section~\ref{sec:disc}, and conclusions drawn in
Section~\ref{sec:conc}.
\section{Characterising detector performance}\label{sec:meat}
\begin{figure*}
\begin{center}
\includegraphics[width=16cm]{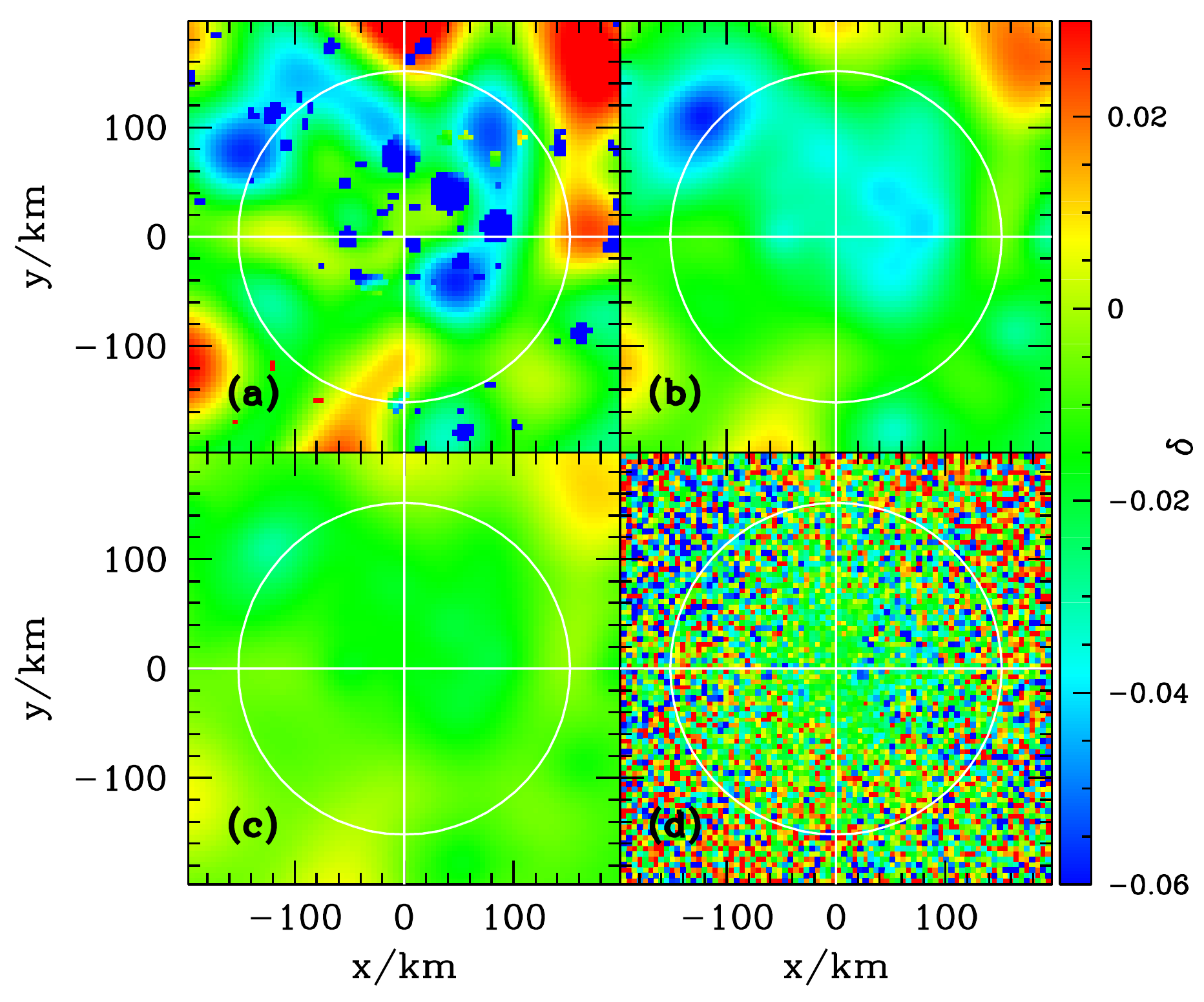}
\end{center}
\caption{Maps illustrating the degradation of information during the process 
of making an observation. Panel (a) shows an example input map
of fractional count rate difference, $\delta$. The map resulting from smoothing with
the spatial response function of the LPNS at $30$ km altitude is shown
in (b). Including a uniform background with the same mean count rate
as that in the input map leads to the results in panel (c), and a
noisy realisation of this, which is what would be measured, is shown
in (d).}
\label{fig:obs}
\end{figure*}
There are three important ways in which measured maps of epithermal
neutron count rate will be degraded representations of what actually
leaves the lunar surface. A detector orbiting above the Moon does not
solely receive neutrons from directly beneath it. Omni-directional
detectors count neutrons coming from all parts of the Moon out to the
horizon, whereas an ideal collimated detector would have a restricted,
but still extended, field of view. In both cases, the measured
epithermal neutron map will be blurred by the extended spatial
response function, or `footprint', of the detector. The blurring caused by the LPNS at
an altitude of $30$ km is illustrated by the difference between the
images in panels (a) and (b) in Fig.~\ref{fig:obs}. 
Rather than
showing a count rate map, these maps show the fractional difference,
\begin{equation}
\delta({\mathbf x})=\frac{c({\mathbf x})-\bar{c}}{\bar{c}},
\end{equation}
where $c({\mathbf x})$ is the count rate in the two-dimensional map
at position ${\mathbf x}$ and $\bar{c}$ represents the mean count
rate per pixel in the observed region. This statistic is invariant
under changes in detector efficiency or cosmic ray flux and thus
represents a convenient way to compare detectors. Panel (a) is a
pixon reconstruction of the LPNS data in the south polar region
\citep{luis10}, with the south pole at the centre of the image, and
adopting a polar stereography projection.
The mean count rate, $\bar{c}$, is defined in the region $|x|,|y|<600$ km.

The second important degradation introduced during the measurement
process arises from the production of neutrons local to the
detector due to cosmic rays striking the spacecraft itself. The
resulting neutrons have nothing to do with the lunar surface composition 
and provide a uniform spatial background that dilutes the contrast
present in the lunar signal, as shown by the difference between panels
(b) and (c) in Fig.~\ref{fig:obs}. A background count rate equal
to the mean in the blurred image has been assumed. This is more 
appropriate for the LEND CSETN than the LPNS
because, while the LPNS was on a boom $2.5$ m away from the main body of
a relatively small spacecraft, the LEND is right next to the much
more massive LRO. This uniform background is distinct from the
background due to the uncollimated lunar neutrons that are scattered
off spacecraft material into the LEND CSETN detector and provide a
spatially varying background. 

The final aspect of the measurement procedure that acts to obscure the
underlying lunar signal is the fact that integration times are finite,
leading to inevitable stochastic noise in the collected
data. Panel (d) of Fig.~\ref{fig:obs} shows how this noise impacts upon the
fractional count rate difference for a sampling similar to that made
by LP. The fact that pixels near to the pole receive more visits and
suffer less statistical noise is clearly visible in this map.

It is evident from Fig.~\ref{fig:obs} that all three of these aspects
of detector performance leave strong imprints on the measured data
set. Thus, determining the relative merits of the LPNS and LEND CSETN
boils down to choosing appropriate statistical measures that are
sensitive to each of these contributing factors. In this way the size
of the instrumental spatial footprint, the background contamination,
and the statistical noise can be estimated empirically from the maps
constructed using data from these two experiments.

Two powerful statistical measures that are widely used in many
different scientific disciplines to quantify the properties of
continuous stochastic fields such as $\delta({\mathbf x})$, 
are the {\it power spectrum} and the {\it autocorrelation function}
\citep{peebles80,monin1statistical}. Both of these quantities
encode information about the amount of structure contained in a map on
a variety of different spatial scales. In the
field of Space Science, these statistical measures have
been used in studies of the lunar gravitational potential
\citep{wei98}, 
modeling of martian dunes dynamics \citep{Narteau:2009JGRF..114.3006N},
helioseismology \citep{dal85}, X-ray
variability from black hole accretion discs \citep{mch06}, galaxy
clustering \citep{cole05,eis05} and the cosmic microwave background
\citep[Planck][]{planck13} to name a few examples.

\subsection{The autocorrelation function}\label{ssec:xir}
The autocorrelation function of a map is a measure of the similarity between
values in pixels at different relative
positions. It is defined by 
\begin{equation}
\xi (r=|\mathbf{r}|) \equiv \left < \delta \left( \mathbf{x} \right) \delta \left( \mathbf{x}-\mathbf{r}\right)\right>
\end{equation} 
where the average is over pixel position $\mathbf{x}$ and isotropy
guarantees that $\xi(r)$ is independent of the direction of the
separation of the pixels, 
$\mathbf{r}$. This function depends not only on the intrinsic
clustering properties of the fractional count rate differences, but
also on both the smoothing length imposed on the data by the instrumental
spatial resolution and the amount of uniform background that is introduced.

\begin{figure}[t]
\begin{center}
\includegraphics[width=0.95\columnwidth]{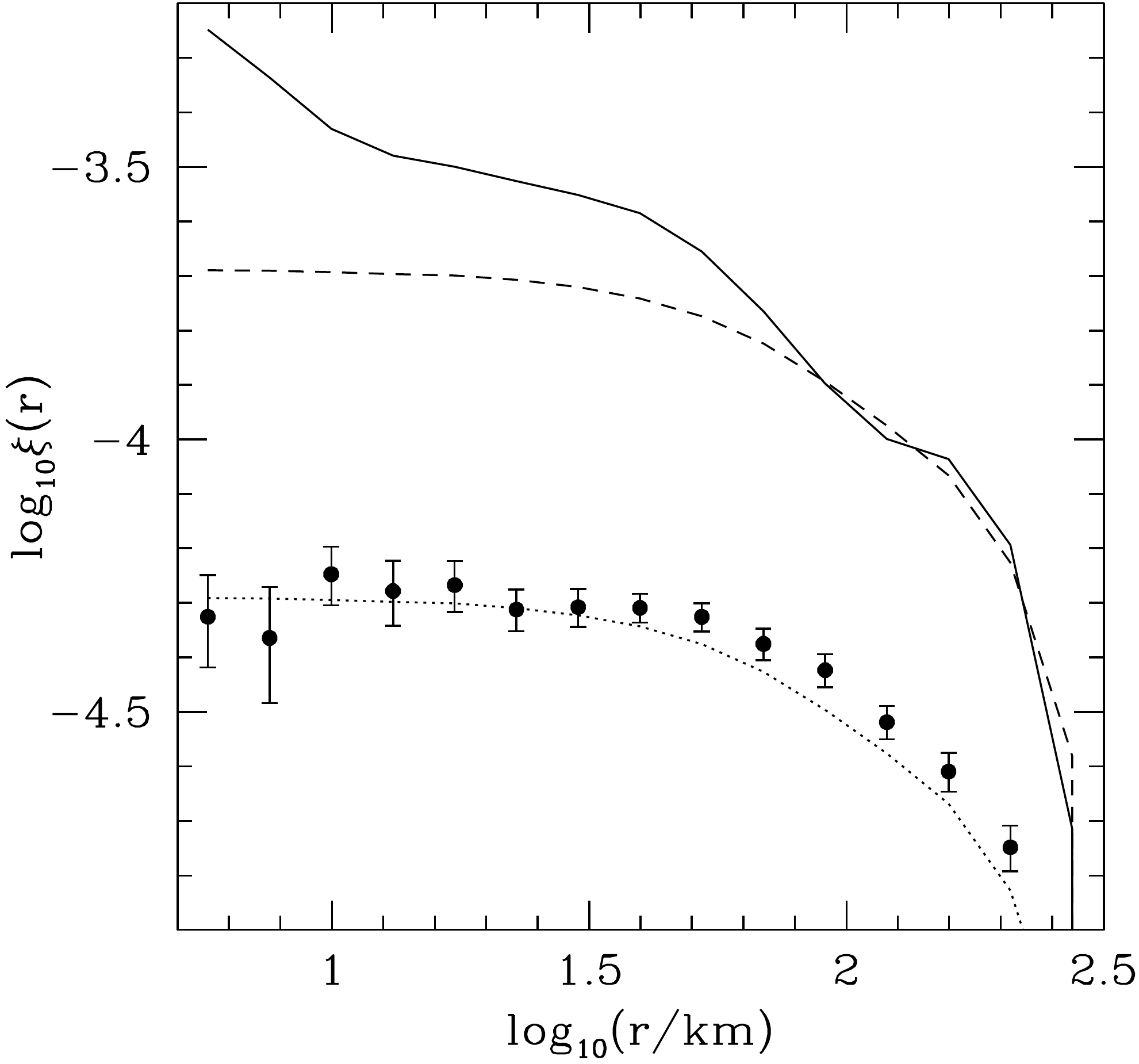}
\end{center}
\caption{The impact of the measurement process on the autocorrelation
function of polar neutron maps. Curves show
the autocorrelation functions for the input map (solid), the blurred
map without any uniform background included (dashed), and after
a uniform background has been added (dotted). These correspond to
panels a, b and c in Fig.~\ref{fig:obs}). Filled circles
represent the autocorrelation function measured from the particular
noisy realisation shown in panel (d) of Fig.~\ref{fig:obs}, with error
bars representing the uncertainty due to sample variance, inferred
using many different noisy realisations.}
\label{fig:egxi}
\end{figure}

In formal terms, smoothing can be represented by the convolution
\begin{equation}
\delta_S(\mathbf{x}) = \int_{\mbox{\scriptsize{over all space }}}
\delta(\mathbf{x}')\, W\left( \mathbf{x} - \mathbf{x}'\right)\, d^2\mathbf{x}',
\end{equation}
where the smoothing kernel is normalised such that 
\begin{equation}
\int W\left(\mathbf{x}-\mathbf{x}'\right)\,d^2 \mathbf{r} = 1,
\end{equation} 
and
$\delta_S(\mathbf{x})$ represents the smoothed map. 
Qualitatively, on scales smaller than the size of the kernel the
correlation function will be approximately flat.
For scales larger than a few smoothing lengths 
the correlation functions of the smoothed and unsmoothed maps
coincide within the measurement error. This is illustrated in
Fig.~\ref{fig:egxi}, where the solid and dashed lines represent
$\xi(r)$ for the unsmoothed and smoothed maps shown in
Fig.~\ref{fig:obs} panels (a) and (b) respectively. The smoothing
kernel used was
\begin{equation}
W(r)=\frac{A}{\left(1+\left(r/\sigma\right)^2\right)^2},
\end{equation}
with $\sigma=35$ km to mimic the omni-directional LPNS at $30$ km
altitude and $A$ being a normalisation constant \citep{maur04}. These
autocorrelation functions have been calculated for polar data on the
projection grid going out to $|x|,|y|=600$ km from the south pole in $5$ km
square pixels (Fig.~\ref{fig:obs} shows the central ninth of this
region), using a similar length of zero-padding to avoid 
wrap-around issues when using Fast Fourier Transforms in the
computation.

The dotted line in Fig.~\ref{fig:egxi} represents the effect of
a uniform background with a count rate equal to that of the mean lunar
signal. This amount of background is far larger than was suffered by the
LPNS, but almost matches that experienced by the LEND CSETN. The
fluctuations in the smoothed map become diluted by a factor of $2$,
meaning that the autocorrelation function is suppressed on all scales
by a factor of $4$. Panel (c) of Fig.~\ref{fig:obs} shows the
associated loss of contrast in the map. $\xi(r)$ for the noisy map in
panel (d) of Fig.~\ref{fig:obs} is represented by the points in
Fig.~\ref{fig:egxi}. As the noise is assumed to be spatially
uncorrelated, only the value of $\xi$ at zero separation (not shown on
this log plot) is systematically changed by the presence of noise. The
larger separation values merely have statistical noise added to
them. These are represented by the error bars, which are determined
from the scatter between the individual measurements when many different
noisy realisations of the same underlying map are made.

\subsection{The power spectrum}\label{ssec:pk}
The power spectrum is just the Fourier transform of the
autocorrelation function and represents an alternative way of showing
which spatial scales contain information. In terms of the wavenumber
$\mathbf{k}=2\pi/\mathbf{\lambda}$, where $\mathbf{\lambda}$
represents the corresponding wavelength in two-dimensional pixel
space, the power spectrum is
\begin{equation}
P(k=\left |\mathbf{k}\right |)\equiv \left < \left | \mathbf{\delta_k} \right |^2 \right >,
\end{equation}
with $\mathbf{\delta_k}$ representing the amplitude of the
$\mathbf{k}$th mode in the Fourier decomposition of the map of
$\delta(\mathbf{x})$ and the average is over modes with the same
wavenumber $k$. 

\begin{figure}
\begin{center}
\includegraphics[width=0.95\columnwidth]{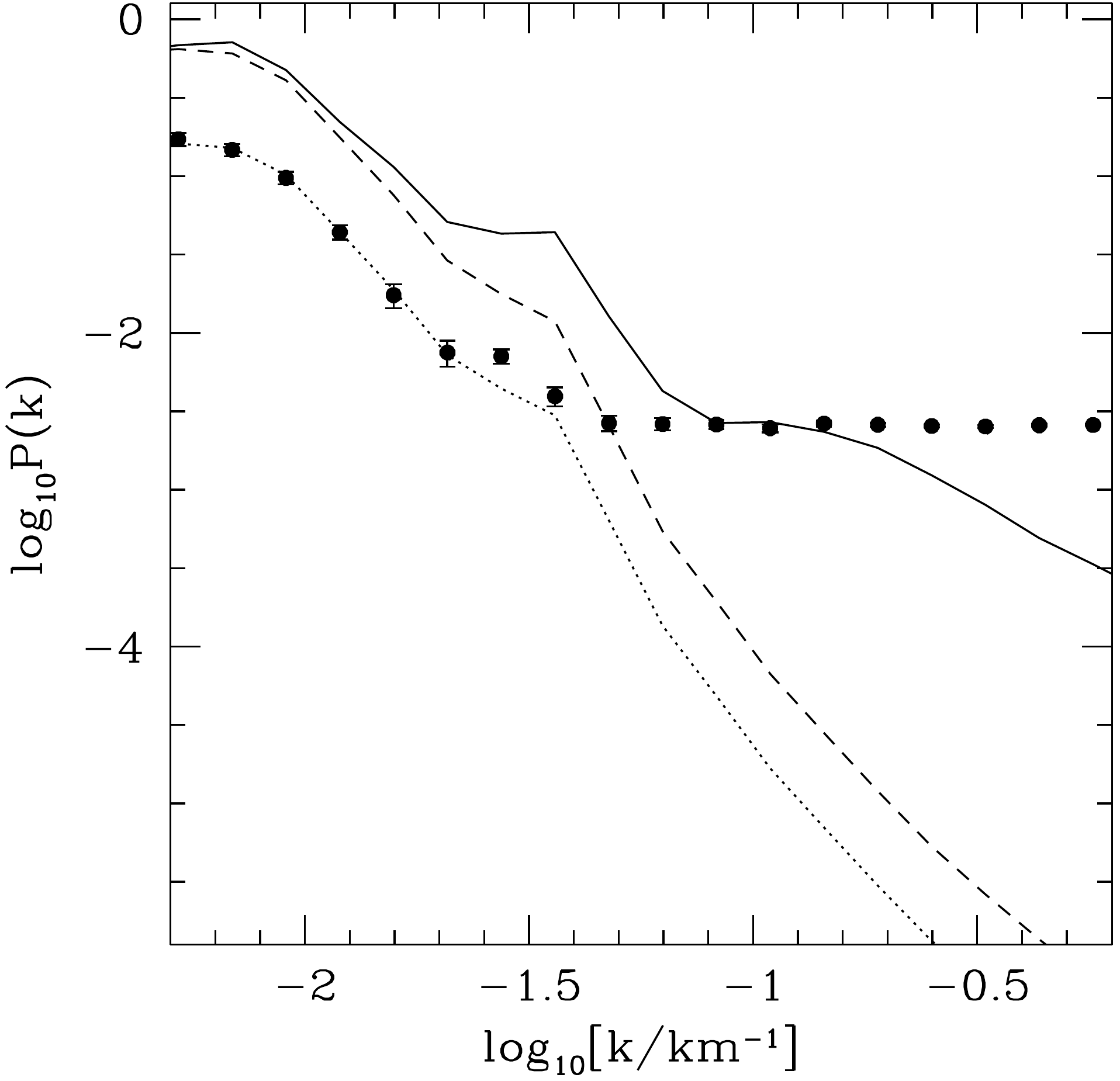}
\end{center}
\caption{The impact of the measurement process on the power spectrum
of polar neutron maps. Solid, dashed and dotted lines show
the power spectra for the input map and blurred maps, with
and without a uniform background added respectively. Filled circles
represent the power spectrum of the particular
noisy realisation shown in panel (d) of Fig.~\ref{fig:obs}, with error
bars representing the uncertainty due to sample variance, inferred
using many different noisy realisations.}
\label{fig:egpow}
\end{figure}

The power spectra for the four maps in Fig.~\ref{fig:obs}
are given in Fig.~\ref{fig:egpow}. The removal of
power at small scales resulting from blurring with the instrumental
footprint manifests itself at large wavenumbers. Once again the
uniform background produces a scale-independent reduction of the power
by a factor of four. However, unlike the case for the autocorrelation
function, where the stochastic measurement noise was confined to the
zero separation signal at $\xi(0)$, this delta function transforms to
a constant in wavenumber space. Consequently, at small scales (large
$k$) where the noise overwhelms what remains of the fluctuations in
fractional count rate difference, the power spectrum goes flat,
identifying precisely the level of statistical noise in the map.

The maps from the LPNS or LEND CSETN data sets can be considered as
the result of the measurement process described in this section.
Both the autocorrelation function and power spectrum of the resulting
maps will provide complementary and comprehensive views of the impact
that the detectors have had on the intrinsic lunar count rate map. In
the following section, these two statistical estimators will be
employed to quantify how well we know the lunar hydrogen distribution. 

\section{Results for lunar neutron data sets}\label{sec:app}
Data from the Geosciences Node of NASA's Planetary Data System 
(PDS\footnote{http://pds-geosciences.wustl.edu}) were used to create
epithermal neutron maps from both the LPNS and LEND CSETN
experiments. The time series Reduced Data Records for the LEND CSETN
were processed almost as described by \cite{boy12}, with a few notable
exceptions. Table 2 of that paper describes the impact that the
various cuts on the data have for the number of one-second data
records that form part of the analysis. However, the quoted number of
total raw 
records exceeds the number of seconds during the claimed period. Thus,
this is impossible to replicate. A couple of other
differences in the reduction procedure adopted here are that an extra
factor of $A_{i,j}$ has been included on the denominator of both equations
(2) and (7), $A_{i,j}$ being the count rate normalisation of the $i$th
sensor during the $j$th switch-on period. Without this extra
factor these equations are dimensionally incorrect. One 
additional important part of the \cite{boy12} data reduction
procedure, not detailed in that paper, is how variances are calculated
for time series records where a subset of the four sensors are working
and they happen to record zero counts. Equation (9) of~\cite{boy12}
appears to suggest that this involves the ratio $(0/0)^2$, which is
not defined. The following equation has been implemented
here to use individual sensor normalisations $A_i^0$ to convert the
individual variances $\sigma_{i,j}^2$ to that on the total `adjusted
count rate', $R$, via 
\begin{equation}
\sigma_R^2=\sum_{on}\sigma_{i,j}^2\left(\frac{\sum_{all}A_i^0}{\sum_{on}A_i^0}\right)^2.
\end{equation}
Presumably \cite{boy12} performed a similar procedure. Other than
these apparent modifications, the treatment of Solar Energetic
Particle events, outlier events, off-nadir measurements, instrumental
warm-up and cosmic ray variation has followed the procedure outlined
by \cite{boy12}. Reassuringly, the results shown here are very similar
to those found using the alternative, independent analysis pipeline
introduced by \cite{eke12}.

\begin{figure}
\begin{center}
\includegraphics[width=0.95\columnwidth]{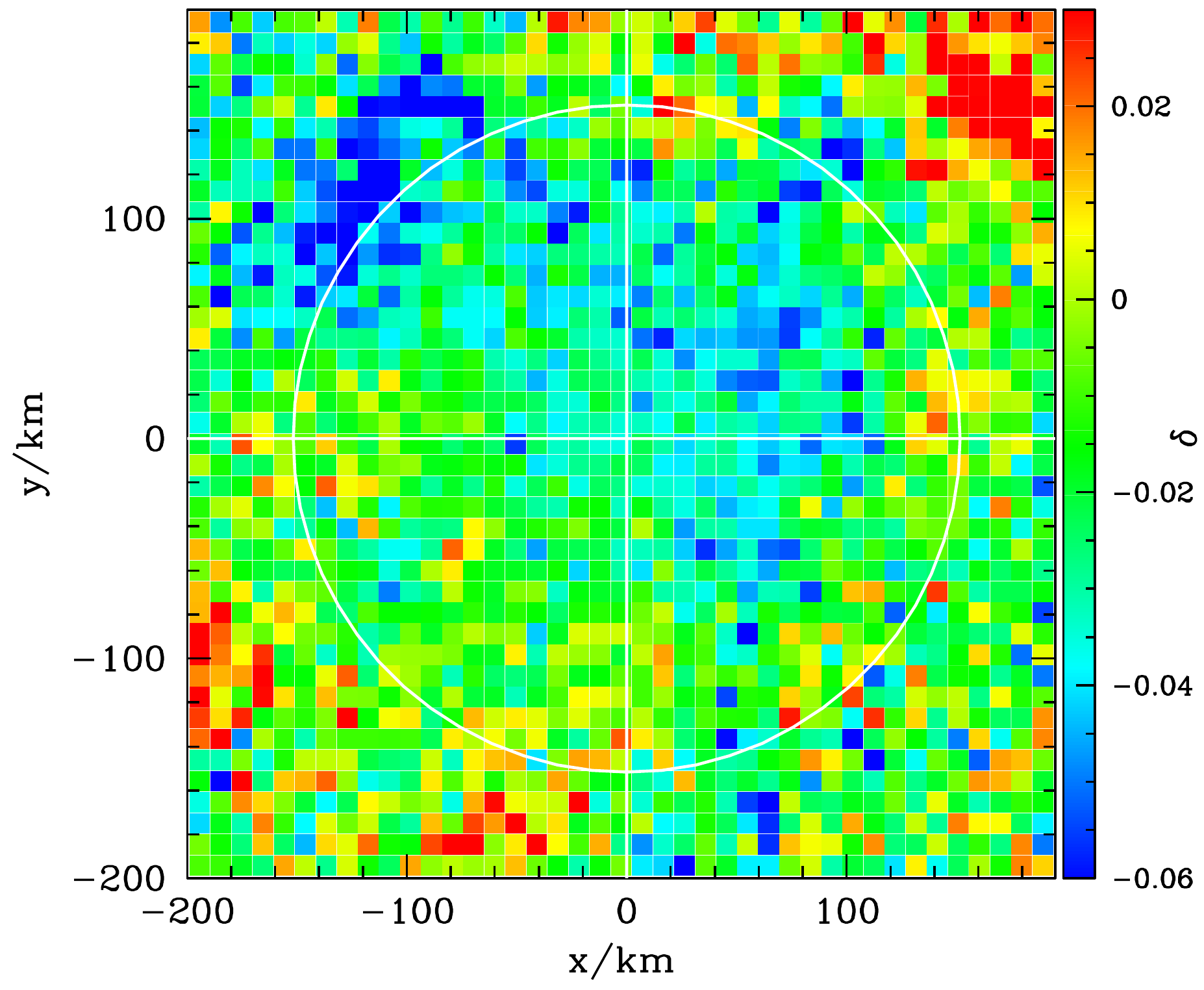}
\end{center}
\caption{Map of the south pole LPNS fractional count rate difference
using $10$ km pixels and data from the $30$ km orbital altitude
period. The white circle represents a latitude of
$-85^\circ$. Statistical uncertainties on the values are $\sim 0.01$
at $-88^\circ$, increasing to $\sim 0.015$ at $-85^\circ$S. Note
that the LPNS data on the PDS has had a small, $\sim 7\%$
\citep{maur04} uniform background component removed.}
\label{fig:lpdelta}
\end{figure}

\begin{figure}
\begin{center}
\includegraphics[width=0.95\columnwidth]{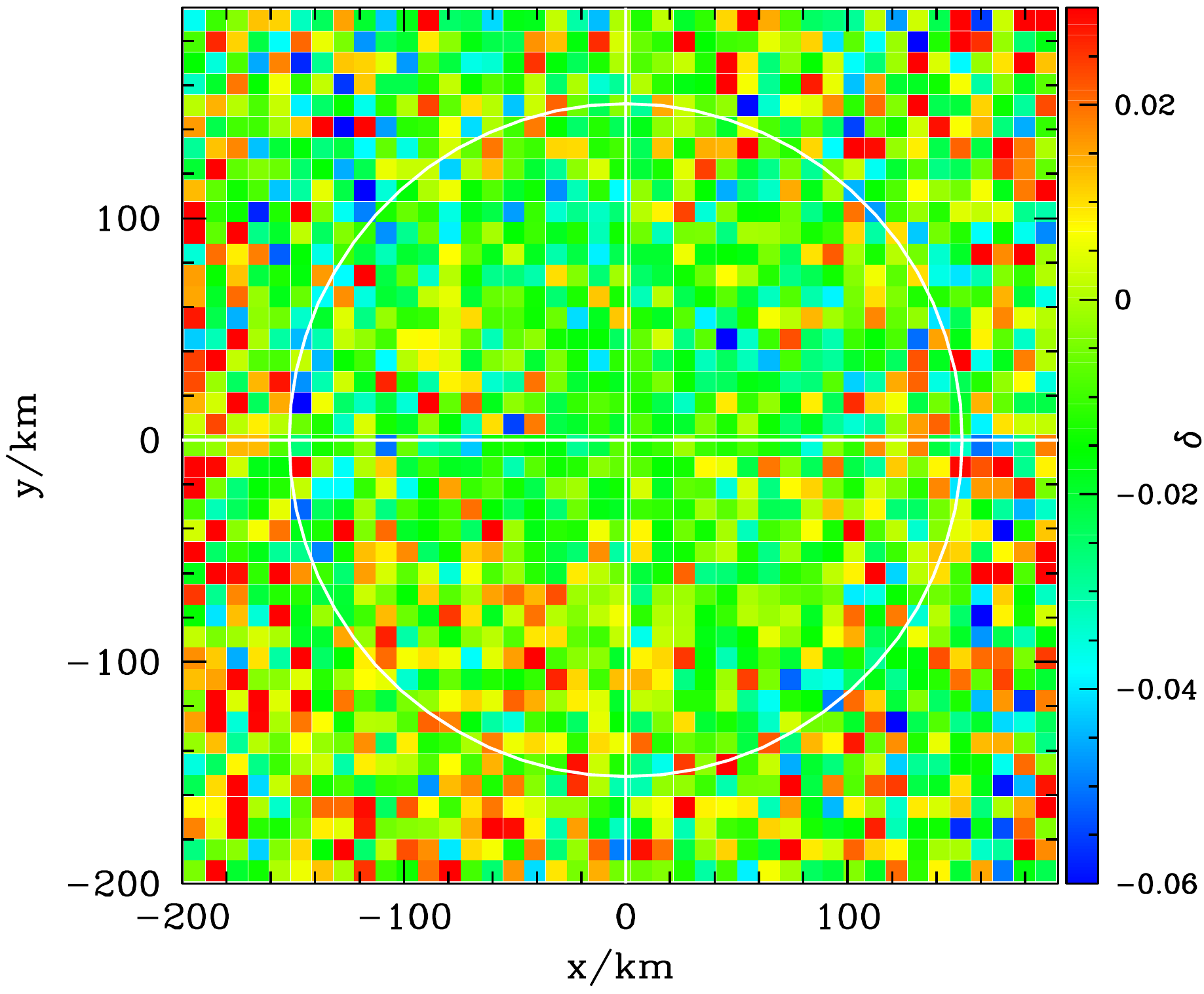}
\end{center}
\caption{Map of the south pole LEND CSETN fractional count rate difference
using $10$ km pixels. The white circle represents a latitude of
$-85^\circ$. Unlike for the LPNS data on the PDS, the LEND CSETN have
not had any uniform background removed. Statistical uncertainties on
the values are $\sim 0.01$ at the pole, increasing to $\sim 0.02$ at
$-85^\circ$.} 
\label{fig:ldelta}
\end{figure}

\begin{figure}
\begin{center}
\includegraphics[width=0.95\columnwidth]{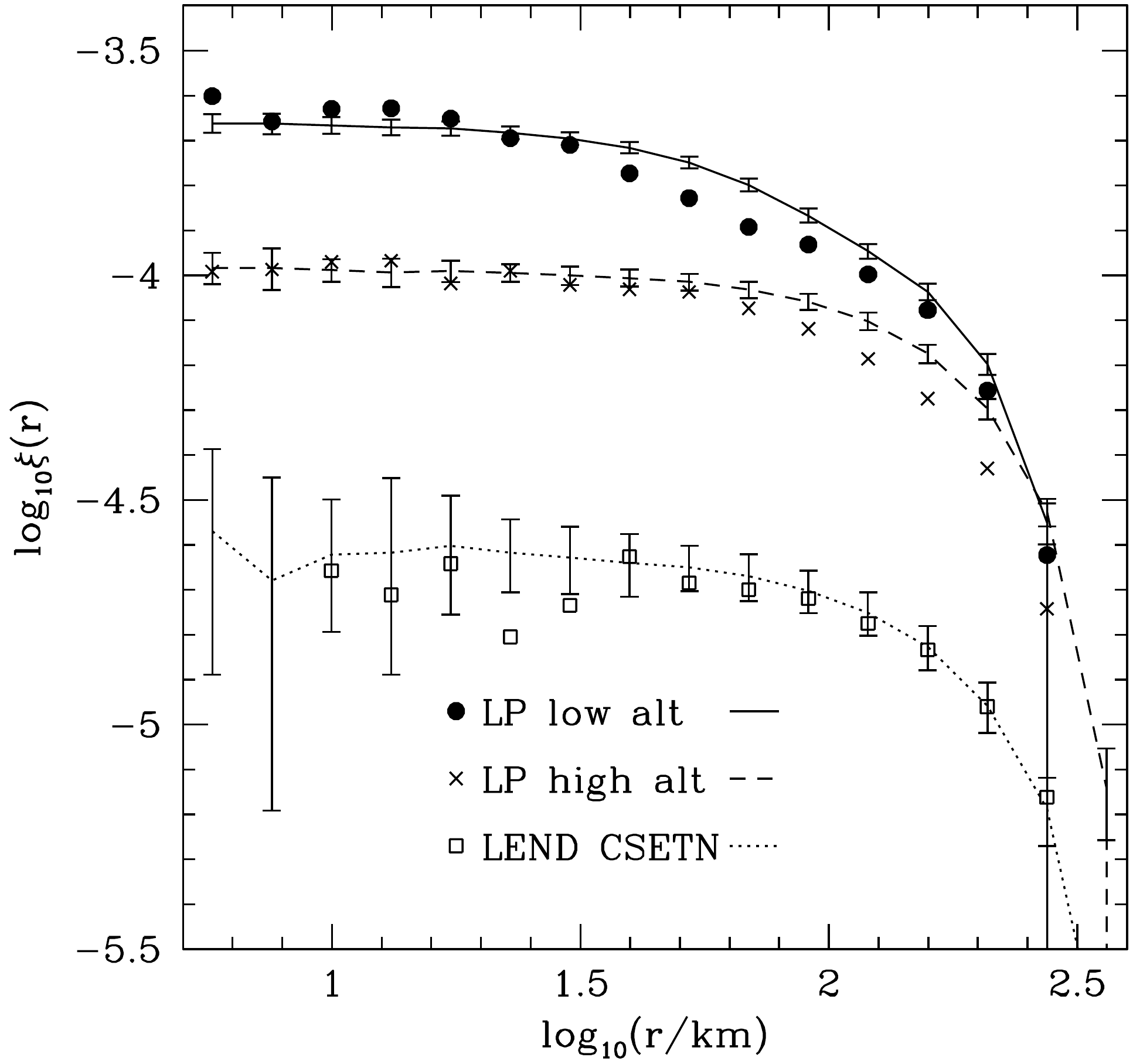}
\end{center}
\caption{Autocorrelation functions for the different experiments. LPNS
results for low and high altitude are shown with filled circles and
crosses respectively. The LEND CSETN results are shown with open
squares. The curves show the mean
autocorrelation functions from 100 different realisations of the
pixon reconstruction of the lunar south pole
\citep{luis10}, and the error bars show the scatter among these
realisations. For the low altitude LPNS mock observations, a smoothing
kernel with $\sigma=35$ km has been used (solid line). The dashed line
assumes that $\sigma=101$ km as is appropriate for the high altitude
LPNS \citep{maur04}. To create the dotted line, a detector with
$\sigma=90$ km, a uniform background fraction of $0.54$ and a
collimated detector fraction of $0.01$ has been used.}
\label{fig:dataxi}
\end{figure}

Figures~\ref{fig:lpdelta} and~\ref{fig:ldelta} show maps of the
fractional count rate difference in the vicinity of the lunar south
pole made using low-altitude LPNS epithermal neutron (seven months) and LEND CSETN ($\sim$ 21 months) data respectively. It should be noted that the $7\%$ uniform spacecraft
background has been removed from the LPNS map \citep{maur04}, whereas the large
uniform spacecraft background is still present in the LEND CSETN PDS
data. 
The pixels used are $10$ km on a
side. The Cabeus region is clearly seen in the LPNS map as the area of
relatively low count rate just over $5^\circ$ from the pole in the
upper left part of the map. This is very much less pronounced in the
LEND CSETN map, which does however have a single $10 \times 10$ km$^2$
pixel with count rate depressed by at least $6\%$ in the Shoemaker
crater on the line $y=x$ at latitude $-88^\circ$. The comparison
between these two maps, made using all available data in the same region
with the same pixellation and the recommended data reduction
procedures for the two different experiments, already makes clear that
the LPNS produces a map with significantly more contrast than the LEND
CSETN. Removing an appropriate uniform spacecraft background from the
LEND CSETN map enhances the contrast present in the map, but also
makes the map look much noisier.
Given the clear 
differences in the information contents present in
the maps for the two different detectors, it is of interest to apply
the statistical estimators described in the previous section to determine
what can be learned about 
these results. Figure~\ref{fig:dataxi} shows the autocorrelation functions
for LPNS data, from both high ($100$ km) and low ($30$ km) altitude
periods and that from the LEND CSETN at its altitude of $50$
km. $5$ km square pixels in a
region out to $|x|,|y|<600$ km from the pole are used. This large area
improves the 
statistical uncertainties, but the conclusions do not change when only
the central ninth of that region (i.e. $|x|,|y|<200$ km), shown in
Figs.~\ref{fig:lpdelta} 
and~\ref{fig:ldelta} with $10$ km square pixels, is chosen. Even
if the uniform spacecraft background comprised a fraction 
$f_{\rm b}=0.54$ of the LEND CSETN count rate, as advocated by \cite{eke12} and
taking into account the fact that this polar region is slightly
different from the entire surface value of $f_{\rm b}=0.535$,
and the autocorrelation function were boosted by a factor 
$1/(1-f_{\rm b})^2$, then it would still lie significantly below that from the
LPNS at low altitude. It would however increase to have a signal
comparable with the high altitude LPNS autocorrelation function. This
implies that the footprint of the LEND CSETN is similar to that of the
LPNS when it was at an altitude of $100$ km.

The curves in Fig.~\ref{fig:dataxi} are constructed by assuming that
the intrinsic map of the lunar south pole epithermal neutron count
rate is that given by the pixon reconstructions of \cite{luis10}. As
the higher energy neutrons 
detected by the LEND CSETN reflect hydrogen variations in an almost
identical way to the lower energy epithermal neutrons measured by the
LPNS \citep{law11b}, it is reasonable to use this 
map for modelling the polar data from the
LEND CSETN. The intrinsic map is observed, by blurring with a
footprint defined by $\sigma$, adding a uniform background, and 
including stochastic noise based upon the observation times in the
different pixels for the different experiments. One hundred different
random noisy realisations are created and the mean of the results from
these forms the curve. The error bars on the curves show the scatter
between individual realisations. The curves for LPNS assume
$\sigma=35$ km and $101$ km for the low and high altitude cases
respectively, and that there is no unaccounted for uniform spacecraft
background. To fit the LEND CSETN data, the model curve has included a
uniform spacecraft background count rate fraction of $f_{\rm b}=0.54$, and made a
composite detector footprint that includes a collimated fraction with
footprint $10$ km of $f_{\rm c}=0.01$ and the remaining
uncollimated lunar flux is collected with a footprint having
$\sigma=90$ km. This is significantly broader than the footprint of an
omni-directional detector at an orbital altitude of $50$ km.

\begin{figure}
\begin{center}
\includegraphics[width=0.95\columnwidth]{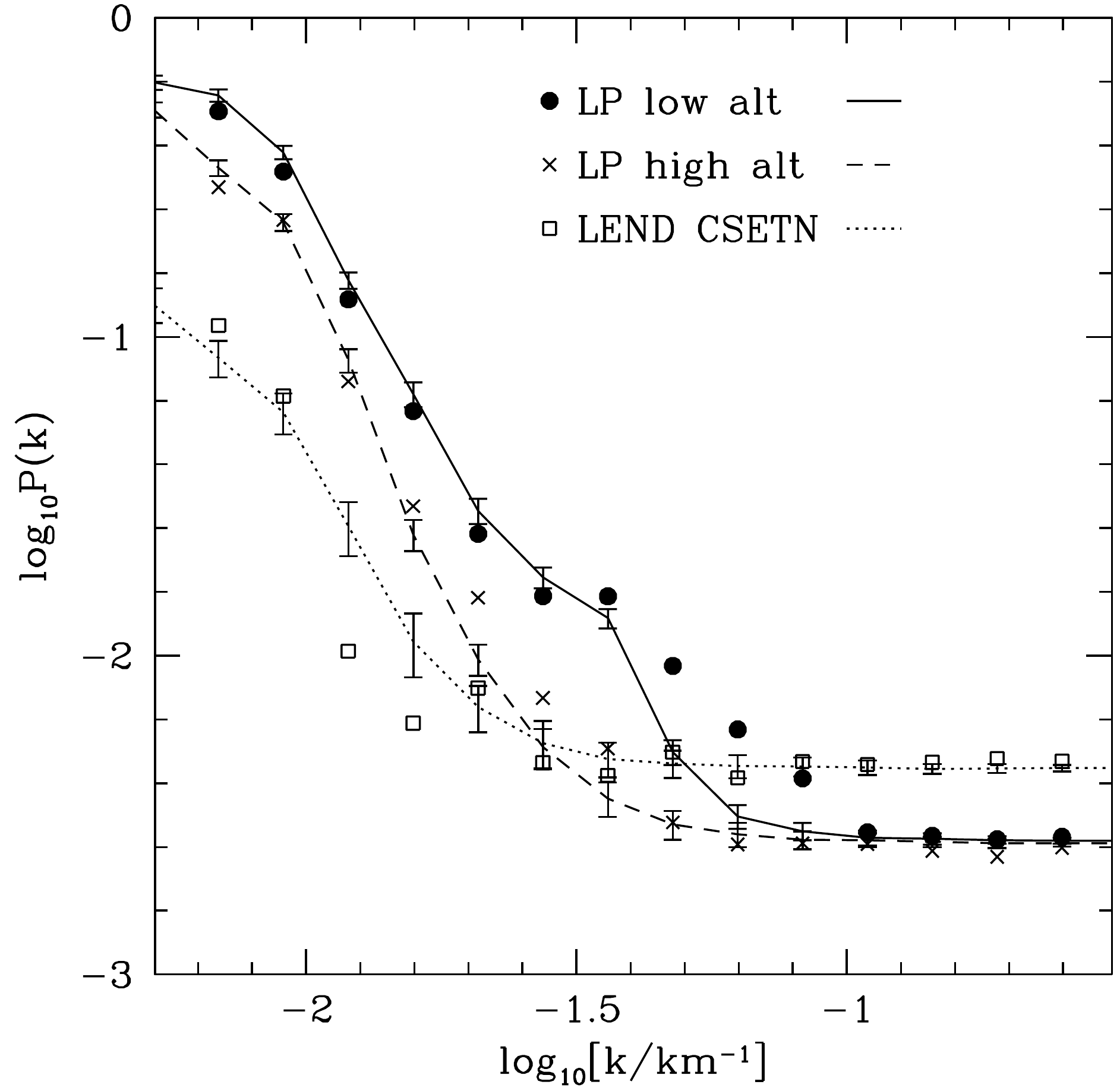}
\end{center}
\caption{Power spectrum results with symbols and curves corresponding
to those shown in Fig.~\ref{fig:dataxi}.}
\label{fig:datapk}
\end{figure}

Figure~\ref{fig:datapk} shows the corresponding power spectra for the
data sets and model fits given in Fig.~\ref{fig:dataxi}. This makes
clear that the noise level at small scales is higher for the LEND
CSETN map than either the low or high altitude LPNS maps. The
low altitude LPNS data only becomes noise dominated for scales smaller
than $\lambda\sim 2\pi/0.1\sim 60$ km. For the high altitude LPNS
data, while the larger footprint suppresses power on intermediate
scales relative to the low altitude case, the noise level is similar
for both data sets. The scale at which noise starts to dominate the
power spectrum of the LEND CSETN map is nearer to
$\log_{10}k=-1.6$ or $\lambda\sim 250$ km. This
should not be a surprise, because the low-altitude LPNS received 
$c_{\rm LP}\sim 20$ neutrons per second and operated for 
$t_{\rm LP}\sim 7$ months, whereas the total
LEND CSETN count rate is approximately $c_{\rm LEND}\sim 4$ neutrons
per second and the data set is only about three times as lengthy, i.e. 
$t_{\rm LEND}/t_{\rm LP}\sim 3$. 
Thus, we expect the noise level to be
higher by 
$\log_{10}[c_{\rm LP}t_{\rm LP}/(c_{\rm LEND}t_{\rm LEND})]\sim 0.22$, 
which is indeed the offset seen at large wave numbers. 
This noise level for the LEND
CSETN is derived including both the uncollimated lunar and spacecraft
background components in $c_{\rm LEND}\sim 4$ per second. If these
neutrons were not included, then the noise fluctuations in 
the lunar collimated component would be higher by a factor of
$1/f_{\rm c}\sim 100$, effectively washing out any fluctuations on all
scales considered here.

\begin{figure}
\begin{center}
\includegraphics[width=0.95\columnwidth]{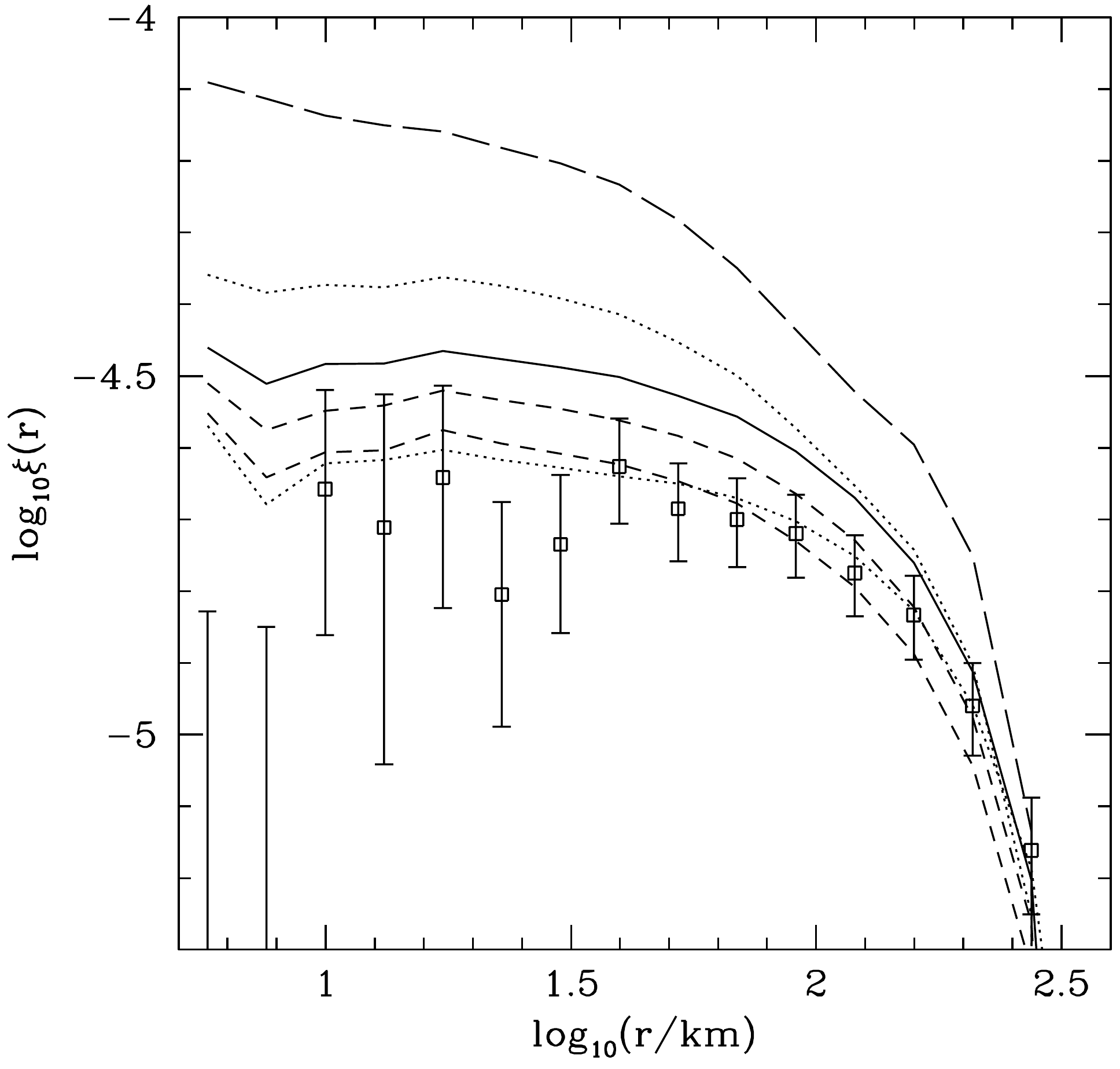}
\end{center}
\caption{Observed autocorrelation function for the LEND CSETN and 
predictions of  various models. 
 The curves result from
observing the pixon south pole reconstruction with various types of
detector. The solid line assumes $\sigma=55$ km, $f_{\rm b}=0.54$ and 
$f_{\rm c}=0.01$. 
The upper and lower dotted lines correspond to 
$\sigma = 35$ and $90$ km, respectively, 
whereas the short-dashed lines
change $f_{\rm b}$ to be $0.57$ and $0.6$. The long-dashed line is the
result of assuming the component fractions advocated by Mitrofanov et
al. (2011),
namely $\sigma=55$ km, $f_{\rm b}=0.46$ and $f_{\rm c}=0.32$.}
\label{fig:lendxi}
\end{figure}

The model fits to the LEND CSETN autocorrelation function and power
spectrum coming from maps of the south pole region suggest that not
only is there a large uniform spacecraft background, but the lunar
neutrons are detected with a footprint that is even wider than would
be expected for an omnidirectional detector at the LRO altitude of
$50$ km. One can now ask how the predicted LEND CSETN autocorrelation
function varies as a function of instrumental footprint and uniform
spacecraft background fraction. Could other combinations also lead to
the measured results.
Figure~\ref{fig:lendxi} shows the autocorrelation function of
the LEND CSETN south pole map with a number of different model curves
included. For clarity, the typical error bars for the models have been
placed onto the data points themselves. The long-dashed curve shows
how a detector receiving a uniform background fraction of 
$f_{\rm b}=0.46$ and a collimated neutron fraction of $f_{\rm c}=0.32$
would perform, under the assumption that the uncollimated lunar flux
was gathered with a footprint having $\sigma=55$ km as would be
expected for an omni-directional detector at an altitude of $50$ km.
These component fractions are the ones reported by \cite{mit11} and
produce a model that clearly does not match the data. As a
consequence, the count rate fractions of \cite{mit11} are ruled out as
a valid description of the composition of the LEND CSETN count rate.

\begin{figure}
\begin{center}
\includegraphics[width=0.95\columnwidth]{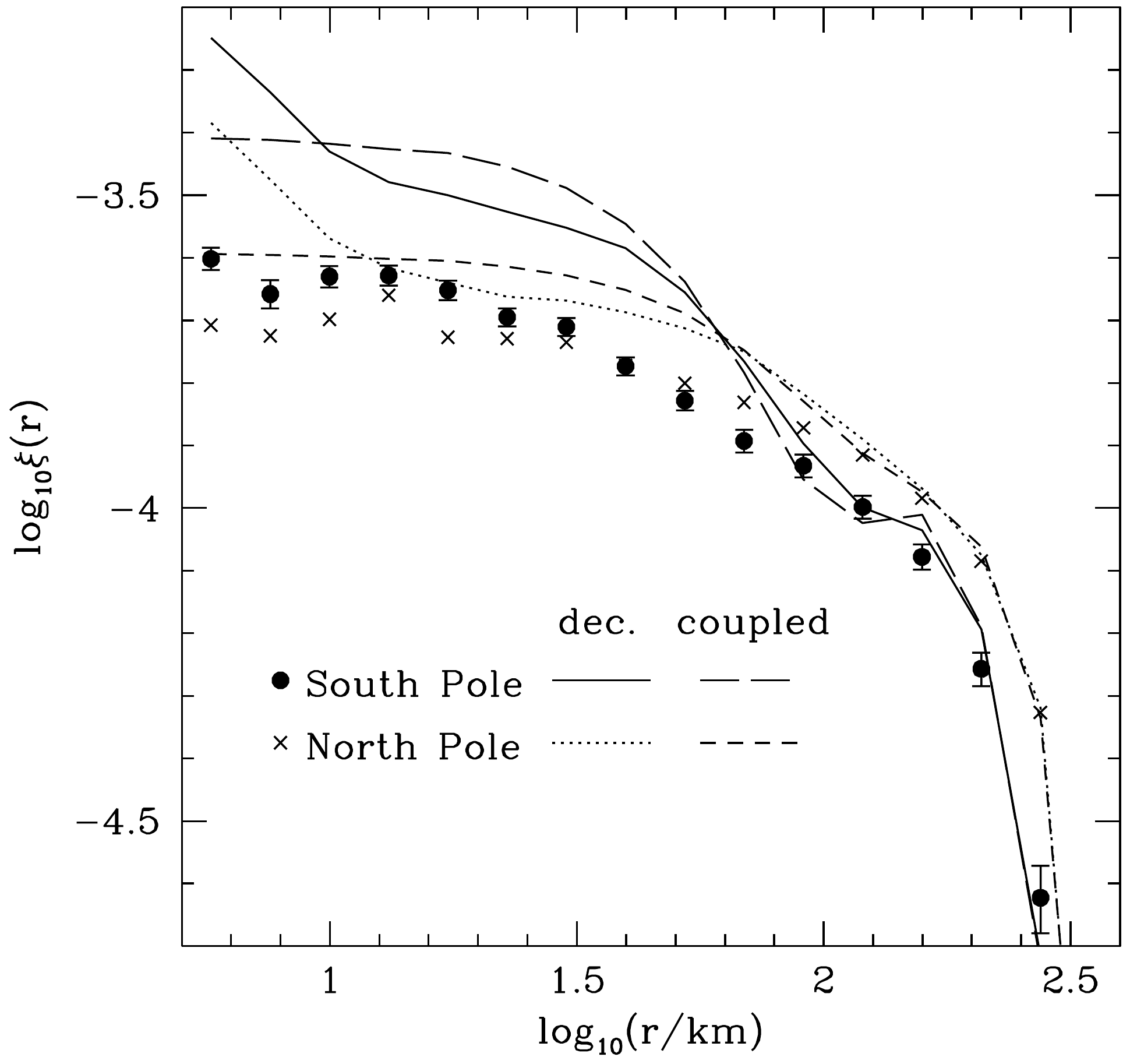}
\end{center}
\caption{Autocorrelation functions for low altitude north (crosses)
and south (filled circles) pole LPNS data and their pixon
reconstructions. Coupled and decoupled reconstructions are shown for
the south pole using long-dashed and solid lines respectively. The
corresponding north polar reconstructions are represented with
short-dashed and dotted lines.}
\label{fig:lpxi}
\end{figure}

A more interesting question is can one really
infer that the footprint is wider than omni-directional, or is it
feasible that a slightly higher uniform spacecraft background can
achieve the same degradation of the signal. The remaining curves in
Fig.~\ref{fig:lendxi} all assume $f_{\rm c}=0.01$ and either vary 
$f_{\rm b}$ from the default value of $0.54$ or $\sigma$ from its
default of $55$ km. The solid line shows the result for these default
values, and it lies systematically above the LEND CSETN data. The two
dashed lines show how increasing $f_{\rm b}$ suppresses the
correlation, with $f_{\rm b}\approx 0.6$ fitting the
data. Despite considering many potential systematic uncertainties,
\cite{eke12} could not produce a set of assumptions that led to 
$f_{\rm b}$ as large as $0.60$. Thus it seems difficult to explain the lack of
contrast in the LEND CSETN data using extra uniform background. The two dotted
curves in Fig.~\ref{fig:lendxi} show the model predictions for
omnidirectional detectors at $30$ km and $90$ km, the latter of which
provides the best fit to the shape and amplitude of the LEND CSETN
results. On the balance of the available evidence, it appears that the
footprint of the LEND CSETN is actually {\it wider} than would be the
case for an omni-directional detector at the altitude of LRO. This is
a possible consequence of the scattering of lunar neutrons off LRO
itself. 
These results make abundantly clear that the LEND CSETN count rate comes
predominantly from the uniform spacecraft background component that carries no
information about the lunar surface. Of the detected neutrons that do
originate from the Moon, the vast majority are {\it not} from the
collimator field of view, but from the uncollimated, spatially-varying
lunar background. It appears likely that the effective
footprint of the LEND CSETN is even larger than would have been expected
for an omni-directional detector at the LRO orbital altitude of $50$
km. While longer integration times will reduce the noise level in the
LEND CSETN map evident at large wavenumbers in Fig.~\ref{fig:datapk},
this will not systematically change the power on larger scales where
the noise does not dominate. The LEND CSETN map, even adjusted for the
uniform spacecraft background, will still remain a diluted version of
the low altitude LPNS map.

Having determined that the low altitude LPNS represents the best data
set for mapping the lunar hydrogen distribution, one can then ask how
do image reconstruction algorithms alter the accessible information
from this data set, and address the question posed in the title of
this paper. Figure~\ref{fig:lpxi} shows results for both the north and
south poles. There are interesting differences between the north and
south pole data sets, with larger signals on $100$ km scales in the
north and the south containing higher contrasts on scales below $30$
km. This presumably reflects the different crater sizes and nature of
the hydrogen distributions in these two regions.

The pixon reconstructions amplify these differences and enhance the
contrast significantly on small scales. Even the coupled
reconstructions of \cite{eke09}, which did not allow the count rates
in the cold traps to vary independently from those in nearby sunlit
regions thus leading to a smooth reconstruction, increase the 
correlation function by a factor of $\sim 2$. The reconstructions that
decoupled the cold trap pixels from the sunlit pixels, allowing larger
contrasts to be found between cold trap and sunlit regions,
show more power on scales less than $\sim 10$ km
than the coupled reconstructions, but only by moving power from $\sim
30$ km scales. As shown by both \cite{eke09} and \cite{luis10}, these
decoupled reconstructions provided better fits to the residuals in the
vicinity of cold traps, and thus represent the best currently
available maps of the lunar polar hydrogen distribution. 

\section{Other evidence}\label{sec:other}
The results in the previous section rather raise the question as to
what is wrong with the arguments put forward by those advocating that the
LEND CSETN is an effective collimated neutron detector. This section
addresses these various claims in more detail.

\subsection{The altitude dependence of the LEND CSETN data}\label{ssec:alt}

One piece of evidence presented by \cite{eke12} that the lunar flux
into the LEND CSETN was predominantly uncollimated was the altitude
dependence of the count rate. The three components contributing to the
neutron count rate should have different variations with detector
altitude. Spacecraft-generated neutrons increase in count rate as the
detector moves away from the Moon and less cosmic ray shielding
occurs. The collimated component should have a rate that is roughly
independent of 
altitude, provided the collimator field of view remains filled by the
lunar disc. In contrast, the count rate of uncollimated neutrons will
decrease as the detector moves to higher altitudes and the Moon subtends
a smaller solid angle. Given that the
majority of detected neutrons in the LEND CSETN are generated from
cosmic rays striking the spacecraft \citep{eke12} and the overall count rate
decreases with increasing altitude of the detector, it is apparent
that there must be a significant lunar uncollimated neutron component,
as quantified by \cite{eke12}.

The recent set of papers claiming that the LEND CSETN has a $10$ km
footprint do not explain the altitude dependence of the observed count rate.
The nearest these authors come to discussing the altitude dependence
is in \cite{lit12b}, who state that
``Data from the commissioning orbit is the important part of the
instrument in-flight calibration because it measured at the variable
altitude above the Moon. ... In this paper we did not discuss these
measurements in details and did not use it as part of the
data reduction process.'' 
In short, the commissioning data provide a valuable way to assess the
performance of the LEND CSETN. Yet, neither \cite{lit12b} nor any of
the other papers in this set report any results from the commissioning
orbit data.

\begin{figure}
\begin{center}
\includegraphics[width=0.95\columnwidth]{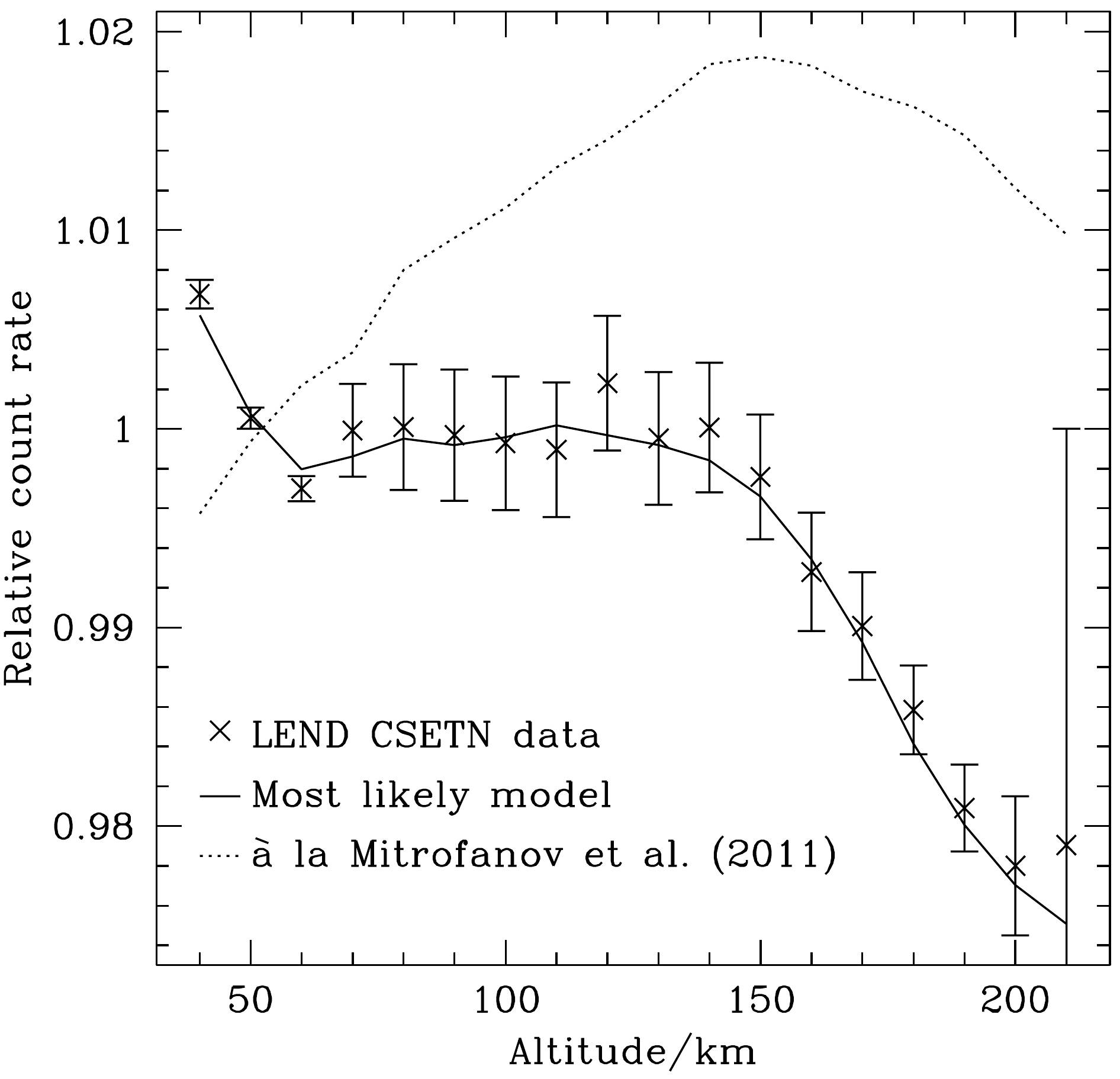}
\end{center}
\caption{Variation of the count rate, relative to the mean throughout
the time series, as a function of altitude. The crosses with error
bars represent the LEND CSETN data, reduced as described by
Eke et al. (2012) and the solid line shows the most likely model fit with
collimated fraction, $f_{\rm c}=0$, a uniform background fraction of
$f_{\rm b}=0.57$ and the remaining $0.43$ in uncollimated lunar higher
energy neutrons. Component fractions as advocated by Mitrofanov et
al. (2011) lead to the dotted line.}
\label{fig:altvar}
\end{figure}

Fortunately, the LEND CSETN commissioning phase data are now publicly
available on the PDS and can be included into the likelihood analysis
presented by \cite{eke12}, who did not have access to them. The
results of performing this experiment 
combining the 80 days of commissioning data with the mapping data from 15th
September, 2009 until the end of 2010, are shown in
Fig.~\ref{fig:altvar}. Count rates are divided by the mean over the
whole time series to give the relative count rate as a function of
altitude. Error bars are much larger for altitudes above $60$ km,
at which the detector was orbiting only during the short commissioning
phase. A reanalysis of the time series in the manner of \cite{eke12},
including the commissioning phase data, leads to most likely component
fractions that are $f_{\rm c}=0.00$, $f_{\rm b}=0.57$ and an
uncollimated lunar count rate comprising a fraction $f_{\rm u}=0.43$
of the total LEND CSETN count rate. These are very similar to those
found by \cite{eke12}, namely $f_{\rm c}=0.01, f_{\rm u}=0.455$ and
$f_{\rm b}=0.535$. This model fits the data well over 
the range of altitudes. The fact that the data do not decrease in a
monotonic fashion with altitude is a consequence of the fact that the
elliptical commissioning orbit had a periapsis over the lunar south
pole \citep{lit12b} and intermediate altitudes were only attained over
equatorial latitudes.
 This is where the iron-rich mare produce a higher flux of energetic neutrons, owing to the higher average atomic mass there.  This effect in CSETN can be seen in figure~10 of \cite{lit12a}.
Only the north pole is measured at
altitudes of $\sim 200$ km, so it is a combination of the lower
intrinsic count rate and the altitude dependence of the components
that leads to the rapid drop off in count rate at high altitude. The
dotted line shows the model with the count rate component fractions
advocated by \cite{mit11}. For the highest altitudes, the
decrease in count rate over the north pole is sufficiently strong that
even this model with a large lunar collimated component produces a
decrease of count rate with increasing altitude. However, when the
relative count rate is fixed at $50$ km, the component fractions
advocated by \cite{mit11} are clearly ruled out by the data at all
other altitudes. This result strongly reinforces those of Section 3
and \cite{eke12} that the component fractions of \cite{mit11} 
are inconsistent with LEND CSETN collimated and background count rates.

\subsection{Shoemaker crater}\label{ssec:shoe}
Shoemaker crater has a diameter of $\sim 50$ km and is located at a
latitude of $-88^\circ$. This crater covers just $\sim 0.02\%$ of the
lunar surface, yet the arguments made by \cite{lit12b} and
\cite{boy12} that the LEND CSETN is producing a high spatial resolution
map rely strongly on data from this location.

\cite{lit12b} argue that, using just over two years of data, the LEND
CSETN provides a $4.6~\sigma$ significance detection of a lower count
rate in the Shoemaker crater relative to the rest of the annulus at
the same latitude, and imply that this provides evidence of the proper
functioning of the collimator. Shoemaker crater
has a diameter of $50$ km, whereas the rest of the annulus at
$-88^\circ$ is $\sim 300$ km long and the collimator field of
view is $10$ km. Repeating the measurement using data from the
omni-directional LPNS reveals a $\sim 4~\sigma$
count rate deficit using only seven months of
low-altitude data. Given that the LPNS was an omni-directional
detector, this demonstrates that Shoemaker crater is too large relative to
the field of view of the LEND CSETN collimator for such a measurement 
to pertain to the effectiveness of the collimator.

The analysis of Shoemaker crater by \cite{boy12} claims that there
is a significant and narrow dip in the measured count rate, which is
much sharper than the broader dip present in the LPNS data. From the
results in the previous section, where it was shown that the LPNS
maps had much greater contrast than those from the LEND CSETN, even
when the uniform background contribution was removed from the LEND
CSETN map, one should immediately suspect that any sharp dips
must be the result of stochastic noise. If the LEND CSETN
did have resolution on $10$ km scales, then the map it produced,
after uniform background correction, would have a higher
autocorrelation function on small scales than that from the LPNS, which
is not the case.

\begin{figure}
\begin{center}
\includegraphics[width=0.95\columnwidth]{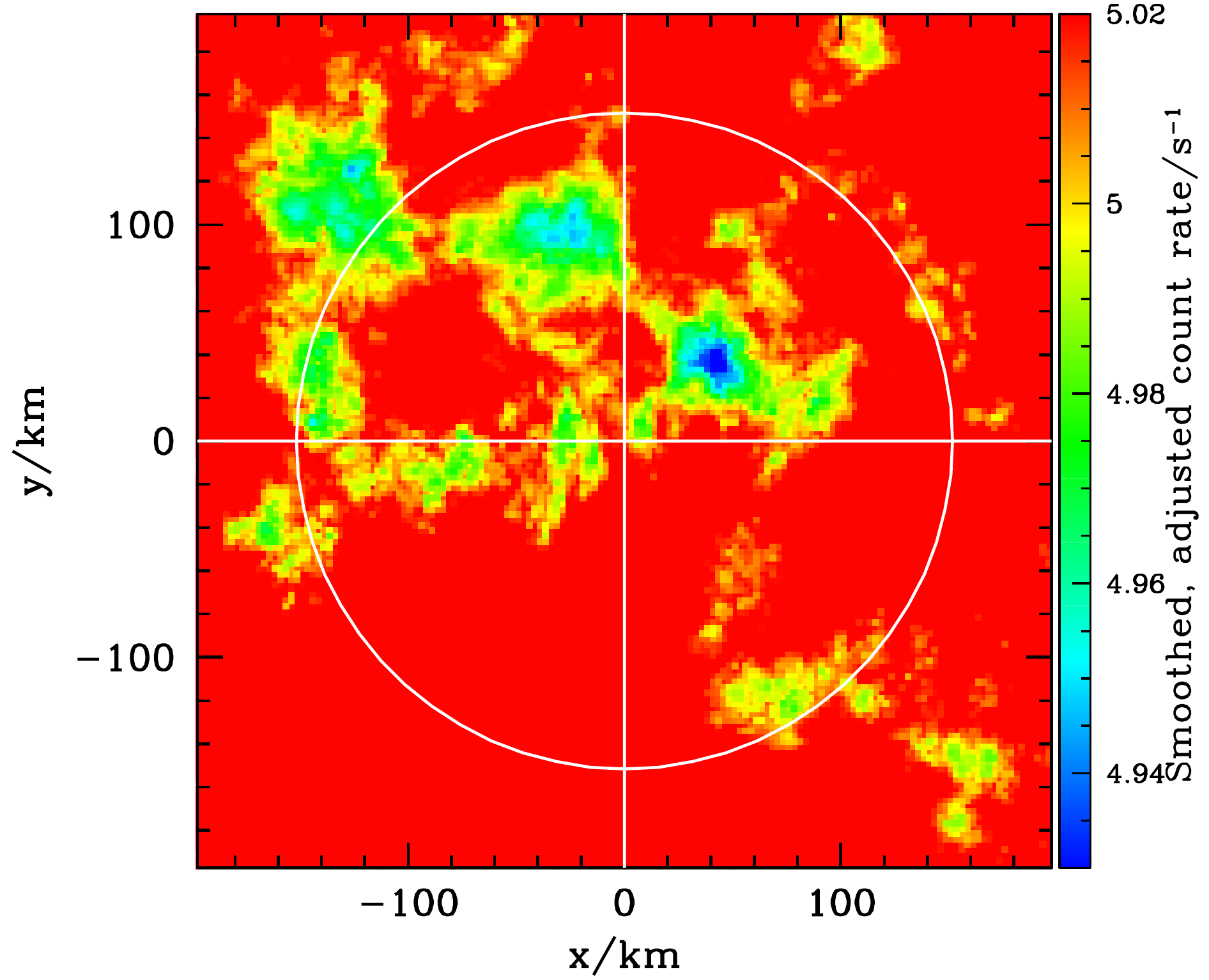}
\end{center}
\caption{LEND CSETN smoothed count rate map of the lunar south pole.}
\label{fig:boyn}
\end{figure}

To try and illustrate that the count rate dip in Shoemaker supports
the claim that the LEND CSETN has a $10$ km spatial footprint,
\cite{boy12} use a latitude-dependent box smoothing with a radius that
is $10$ km at the pole and already $\sim 19$ km at the latitude of
Shoemaker. Adopting this same smoothing of the weighted, adjusted
count rates for the LEND CSETN data reduced as described by
\cite{boy12} with the additional points noted in
section~\ref{sec:app}, leads to the smoothed count rate distribution
shown in Fig.~\ref{fig:boyn}. The mean count rate in the region shown
is $5.04$ neutrons per second, so the colour scale has been
truncated at the high end to try and reproduce figure 7
in \cite{boy12}. While Fig.~\ref{fig:boyn} is quite similar in
appearance to figure 7 of \cite{boy12},
the depth of the depressions in count rate is only about half that
found by \cite{boy12}. Why this is so is not clear. However,
confidence in Fig.~\ref{fig:boyn} can be taken from the fact that it is very
similar to the map found with the independent data reduction process
used by \cite{eke12}. It is also consistent with the simple binning of
the fractional count rate difference in Fig.~\ref{fig:ldelta}. While
there is one pixel in that figure at $(45,45)$ km from the pole within
Shoemaker having $\delta\sim -0.06$, averaging over the region
corresponding to the box-smoothing scale used for Fig.~\ref{fig:boyn}
gives $\bar{\delta}\sim -0.026$. These $\delta$ values are measured
relative to the mean count rate in a region extending $400$ km from
the pole, where $\bar{c}=5.08$. Thus, $\bar{\delta}\sim -0.026$
corresponds to a count rate of $\sim 4.94$ neutrons per second,
consistent with that found in Fig.~\ref{fig:boyn}.

\begin{figure}
\begin{center}
\includegraphics[width=0.95\columnwidth]{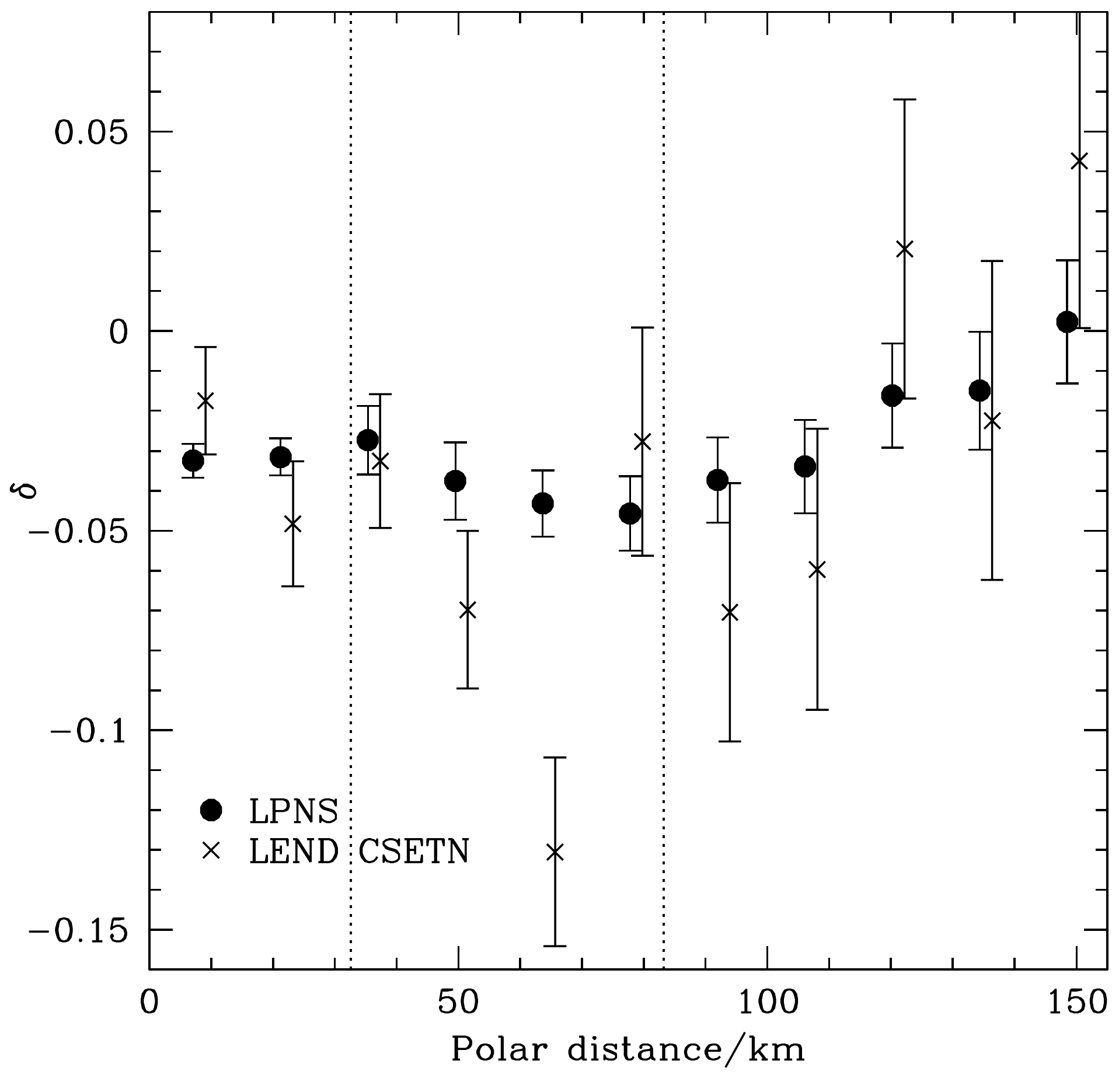}
\end{center}
\caption{Fractional count rate difference as a function of distance
from the south pole along longitude $45^\circ$, through Shoemaker
crater. The filled circles and crosses show results for the LPNS and
LEND CSETN respectively. Each data point corresponds to one of the $10$ km
pixels along $y=x$ in Figures~\ref{fig:lpdelta} and~\ref{fig:ldelta}. A
uniform background of $f_{\rm b}=0.54$ has been removed to enhance
the contrast in the LEND CSETN map. Error bars show the $1 \sigma$
error on the mean $\delta$ in each pixel. The LEND CSETN 
results have been displaced by $2$ km in polar distance for clarity.
Vertical dotted lines delineate the limits of Shoemaker crater.}
\label{fig:trace}
\end{figure}

\begin{figure}
\begin{center}
\includegraphics[width=0.95\columnwidth]{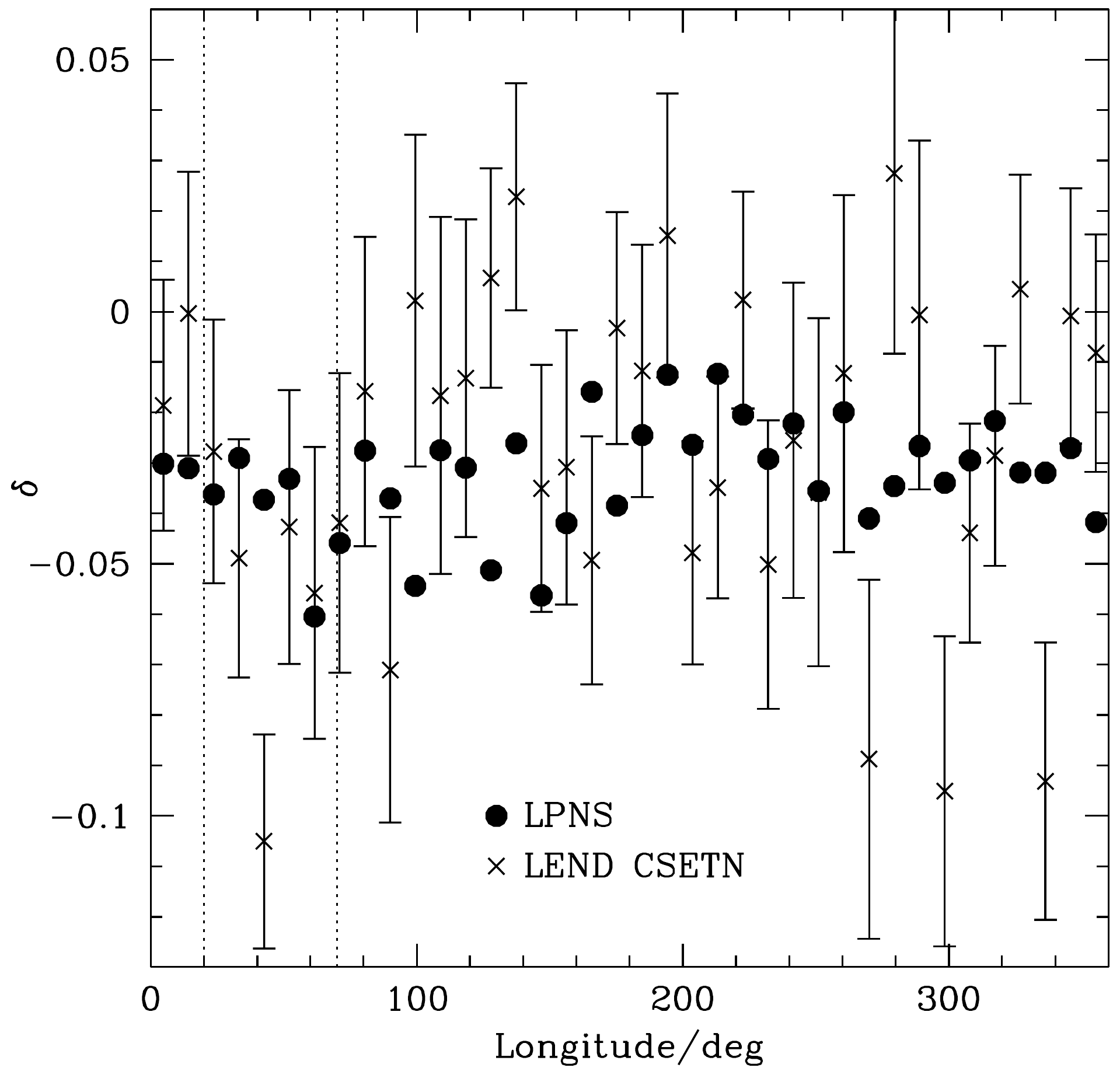}
\end{center}
\caption{Fractional count rate difference as a function of longitude
for a $10$ km-wide annulus at latitude $-88^\circ$.
Filled circles and crosses show results for the LPNS and
LEND CSETN respectively. Data points every $\sim 10$ km in azimuth are
shown and a
uniform background of $f_{\rm b}=0.54$ has been removed to enhance
the contrast in the LEND CSETN map. Error bars on the LEND CSETN
points show the $1 \sigma$ error on the mean $\delta$ in each
pixel. For clarity, the relatively small uncertainties ($\sim\pm 0.01$)
on the LPNS results are not shown. Vertical dotted lines delineate the
limits of Shoemaker crater.}
\label{fig:loop}
\end{figure}

A trace of the fractional count rate difference as a function of
distance from the lunar south pole along longitude $45^\circ$ is shown in
Fig.~\ref{fig:trace}. The LEND CSETN contrast has been amplified by a factor
$1/(1-f_{\rm b})$ to `correct' for the dilution from the uniform
spacecraft background. $f_{\rm b}=0.54$ has been used, which more than
doubles the contrast evident in the uncorrected map. The
centre of Shoemaker crater lies $\sim 60$ km from the pole.
One pixel, $\sim 65$ km from the pole, lies below the
neighbouring LEND CSETN pixel values, and a significant dip on $10$ km
scales would be indicative of a significant collimated component of
the LEND CSETN count rate. Thus the question is, how significantly far
beneath the results for neighbouring pixels does this one lie?
The error bars shown on these points only represent the statistical
uncertainties associated with the counting experiment and how they are
altered by the various adjustment and correction factors applied
through the data reduction procedure. They do not include inherent
systematic uncertainties associated with the various
corrections and should thus be viewed as appropriate
for the case of zero systematic errors in the data reduction
procedure. Under this assumption, the significance of the difference
between two pixel values, $\delta_1\pm\sigma_1$ and
$\delta_2\pm\sigma_2$ will be $S$ standard deviations, where
\begin{equation}
S=\frac{|\delta_1-\delta_2|}{\sqrt{\sigma_1^2+\sigma_2^2}}.
\end{equation}
Thus the pixel at $\sim 65$ km is just under $2 \sigma$ below
that at $50$ km and almost $2.8 \sigma$ beneath that $80$ km from the pole.
Given that, averaged over the entire polar region, the LEND
CSETN is less able than the LPNS to detect fluctuations at small
scales, as shown in the previous section, one may safely conclude
that this particular low $10 \times 10$ km$^2$ pixel is entirely
consistent with being a statistical fluctuation. 
One further piece of evidence is shown in Fig.~\ref{fig:loop}, where a
$10$ km wide band around latitude $-88^\circ$ is presented for both
the LPNS and uniform background-corrected LEND CSETN data. The data
points represent $\sim 10$ km-long sections of the annulus, chosen
because this is the field of view of the LEND CSETN collimator. Shoemaker
is situated at longitude $45^\circ$, where a single insignificantly
lower pixel can be seen for the LEND CSETN. Comparably low values of
$\delta$ are seen in three other pixels at higher longitudes. The
larger scatter in the LEND CSETN $\delta$ values, compared with those
from the LPNS, is very clear. If these small-scale fluctuations
represented real features on the lunar surface, then they would show
up as coherent contributions to the autocorrelation function and power
spectrum. They do not. Thus, it is appropriate to conclude that these
small-scale fluctuations are the result of stochastic noise.

\subsection{The lunar uncollimated background}\label{ssec:back}
In order to produce maps of the collimated lunar component,
\cite{boy12}, \cite{san12} and \cite{mit12} remove a component of
uncollimated lunar background
flux from the LEND CSETN count rate. Despite the choice of colour
scheme, figure 9 of \cite{boy12} shows that the range of variation in
combined background from the spacecraft and lunar uncollimated
components amounts to no more 
than $\sim 4$ parts in $1000$. This means that it is essentially
spatially invariant near the pole, and any fluctuations seen in the
total count rate map are ascribed to the lunar collimated
component. This lack of variation in the lunar uncollimated component
at the poles is inconsistent with that predicted by Monte Carlo
neutron transport models \citep{law11b}.
Furthermore, the low value of $1.1$ counts per second assumed by
\cite{boy12} for the lunar uncollimated component would not be able to
recreate the higher count rates seen over mare regions by the LEND
CSETN \citep{law11b,eke12}.

\cite{mit12} adopt a slightly different approach that has a very
similar effect. A ``reference'' map is made by smoothing the LEND
CSETN data on a scale of $\sim 230$ km. The difference between this
greatly smoothed map and one with a latitude-dependent smoothing
radius of $11$ km at the pole growing to $25$ km at $-70^\circ$
latitude is used to determine where ``local suppression/excess spots''
exist. If the lunar uncollimated background results from neutrons
coming from a scale of $\sim 80$ km across on the lunar surface, as
one might anticipate for an omni-directional detector at an altitude
of $50$ km, then this reference map will be too smooth to include any
regional variations due to uncollimated lunar flux. Any such
variations will then be ascribed to the collimated lunar component,
which is anyway already smoothed on scales larger than the collimator
field of view. There is no way that such an analysis can determine
whether or not the LEND CSETN is behaving as a
collimated detector.

\cite{san12} define their ``local background'' for a given crater
using either a region at the same latitude or the LEND CSETN polar map
smoothed on a similar large scale to \cite{mit11}. The similarity
between figure 1 in \cite{san12} and figure 1 in \cite{mit12} suggests that the
same latitude-dependent smoothing of the map has been used in order to
suppress noise on smaller scales. \cite{san12} conclude that there are
three large permanently shaded regions (PSRs) that contain significant
neutron suppressions, while smaller PSRs do not contain significant
deviations from the background count rate. No effort is made to
compare these measured neutron suppressions, which seem completely in
keeping with what one 
would expect if little of the count rate was actually collimated, with
what would be found with the LPNS. As pointed out in
section~\ref{ssec:alt}, a count rate dip above Shoemaker crater is
also very well detected by the LPNS. Thus, no substantive evidence to
support claims about the functioning of the collimator in the LEND
CSETN can be drawn from the paper by \cite{san12}. 

\subsection{Orbital Phase Profiles}\label{ssec:opp}
\cite{lit12a} use the orbital phase profile to conclude that the
collimated count rate into the LEND CSETN is $1.7$ neutrons per second.
The orbital phase profile involves averaging over narrow latitude
bands either on the near or far side of the Moon. The lengths 
of these bands are very much greater than the field of view of the
collimator, so the purpose of the orbital phase profile is to compare
large-scale features in global maps of different energy neutrons in
order to determine the fractions of the total LEND CSETN count rate in
the lunar collimated and lunar uncollimated components. To achieve
this aim, \cite{lit12a} assume that the lunar uncollimated component
has a variation with longitude and latitude that matches that of the
fast neutrons measured by the LEND Sensor for High Energy Neutrons
(SHEN). Not only is this assumption unjustified, but it is also unjustifiable.
Monte Carlo neutron transport simulations by \cite{law11b} suggest that
high energy epithermal (HEE) neutrons are the primary contributor to the
uncollimated lunar count rate, with a smaller portion from fast
neutrons. This is important 
because, while both HEE and fast neutron fluxes are similarly changed
by the increase in mean atomic mass in the mare regions, the HEE
neutrons are much more sensitive to hydrogen near the lunar
poles. Thus, the assumption made by \cite{lit12a} forces the
large-scale polar count rate dips to be ascribed to a significant
collimated component because the variation of the uncollimated
component has been falsely denied the opportunity to contribute to these
polar count rate dips. Had the altitude dependence of the count rate
variation been simultaneously investigated, then the inappropriateness
of this assumption would have been evident.

\section{Discussion}\label{sec:disc}
The results of the autocorrelation function and power spectrum
analyses contained in this paper indisputably show how, even in the
polar regions, 
the maps from the LEND CSETN are lower contrast than those from the
LPNS on a range of scales. While much of this is due to the dominant,
spatially invariant spacecraft-generated neutron background into
the LEND CSETN, the lunar component of the count rate is also seen to
display less small-scale power than is found by the LPNS. This deficit
of small-scale structure exists to such an extent that the LEND CSETN
results are best described by a model where the detector footprint is
even slightly broader than omni-directional for a 
spacecraft at the $50$ km altitude of LRO. The $10$ km spatial
resolution claimed by some authors \citep{mit10b,mit11,mit12,boy12} is
inconsistent with the LEND CSETN data themselves, and
thus, once again, this should be rejected as a viable
hypothesis. Further, claims that hydrogen enhancements are not
co-located with permanently shaded regions \citep{mit10b} and that
hydrogen enhancements are equally likely in shaded and sunlit regions
\citep{mit12} are not supported by the LEND CSETN data.

With just a straightforward scaling argument, one can see
why the LPNS produces a more significant map of the count rate
variations and hence hydrogen distribution. Suppose that the neutron count
rate measured for regolith containing no hydrogen, $s_0$, were
precisely known for both the LPNS and the LEND CSETN. The ratio of
lunar neutron count rate, 
$s$, to $s_0$ sets the local hydrogen abundance. How much longer would
the LEND CSETN need to collect data to receive the same accuracy in
the derived hydrogen abundance as the LPNS, assuming that they
actually have the same sized footprint? If there were not a large
spacecraft background contribution to the LEND CSETN, then it would
receive just $\sim 2$ neutrons per second, which is about a tenth of
the LPNS rate. Thus it would need an observation period ten times as
long as that of the LPNS to determine $s/s_0$ to the same fractional
precision. However, the uniform spacecraft background contains a
comparable variance to that in the lunar signal itself and the
background and spacecraft neutrons are not distinguishable for the
LEND CSETN. This means that, assuming the mean background count rate
were precisely known, an extra factor of two in integration time is
required to recover the same fractional accuracy in the inferred
hydrogen abundance. Consequently, for the LEND CSETN to match the
hydrogen map from 7 months of the LPNS at an altitude of $30$ km
would require $t_{\rm LEND}\sim 20~ t_{\rm LP} \sim 12$
years. Even then, the LEND CSETN map would be lower spatial resolution
because of the broader instrumental footprint.

At the lower orbital altitude of $30$ km, the omni-directional
LPNS map already contains more significant structure than is present
in that from the LEND CSETN. Furthermore, the application of image
reconstruction techniques has been shown to enhance the contrast in
the count rate map by suppressing the inevitable stochastic noise and
undoing some of the 
blurring that is unavoidably associated with the extended detector
footprint. This objective assessment makes clear that the most
accurate maps of the lunar polar hydrogen distribution are those
resulting from LPNS data processed through an image reconstruction
algorithm. 

The arguments put forward by \cite{mit11,lit12a,lit12b} and
\cite{boy12} in support of 
the LEND CSETN functioning well as a collimated neutron detector have
been considered in the previous section. None of them are found to
provide strong evidence to bolster the claims that the majority of the
lunar component into the LEND CSETN is collimated. Their conclusions
appear to result from a mixture of unjustifiable or demonstrably
incorrect assumptions, a misapplication of statistics, or an
unrepeatable data reduction process. In contrast, the wide range of
measurements considered in detail in this paper are all consistent
with the component fractions inferred by \cite{eke12}; namely that the
uniform spacecraft background produces 
just over half of the counts into the LEND CSETN, with the spatially
varying uncollimated lunar background close behind and the collimated
lunar component providing fewer than $5\%$ of the total counts.

Collimating epithermal neutrons is distinctly non-trivial and the
claims that the LEND CSETN is producing maps with a spatial resolution
of $10$ km are extraordinary. ``In science, the burden of proof falls
upon the claimant; and the more extraordinary a claim, the heavier is
the burden of proof demanded'' \citep{tru87}. However, many lines of
evidence reject the hypothesis that this level of spatial resolution
is achieved. An alternative hypothesis, consistent with the available
evidence, was provided by \cite{eke12}. This hypothesis
has passed the further testing performed in this paper, where the different
techniques employed have further quantified the footprint of the LEND CSETN.
It is likely that plans for future missions to the lunar poles
will use maps of the hydrogen distribution, so it is important that
the capabilities of the LEND CSETN are properly appreciated in order
to prevent costly future mistakes in targetting landers.

Given that the performance of the LEND CSETN instrument has been shown
here to be greatly inconsistent with the claims made by various
authors, and has not met its primary objective of mapping the lunar
neutron flux at a spatial resolution of $\sim 10$ km,
one might reasonably ask how these data can best be used.
The detector
has, as was pointed out by \cite{eke12}, made the first map of lunar
neutrons with this particular energy-dependent filter that is
picking out a mixture of high energy epithermal and fast neutrons. While
the map is noisy and suffers from both a large spacecraft background
and a very extended spatial footprint, it is
still a unique resource. In order to extract scientifically useful
results from this instrument, the challenge will be to understand the
neutron transport within LRO well enough to determine which energies
of neutron is the LEND CSETN measuring, and what they reveal about
the composition of the lunar surface.

\section{Conclusions}\label{sec:conc}
The best available maps of polar hydrogen come from the pixon
reconstructions of LP data. These provide estimates of the average
weight percentage of water 
equivalent hydrogen in polar craters that range up to a few per cent
and have a fractional uncertainty of $\sim 30\%$ \citep{luis10}. The
LEND CSETN produces maps containing a dominant 
background from neutrons that arise due to cosmic ray interactions
with the spacecraft. The
effective detector footprint, taking into account both the collimated
lunar and uncollimated background lunar counts may even be broader
than that for an omni-directional detector at $50$ km altitude. The
suppression in the count rate over Shoemaker crater is consistent
with a statistical fluctuation superimposed upon a broad dip in count
rate of the sort that an omni-directional detector such as the LEND
CSETN would measure in this region. Thus, it does not support the claim
that the LEND CSETN is collimated.

The results of this study are relevant to
the proposed ``Fine Resolution'' Epithermal Neutron Detector
\citep[FREND]{mal12} that is scheduled 
to be launched in 2016 on the ExoMars mission.

\section*{Acknowledgments}
L.T. and R.E. acknowledge the support of the LASER and PGG NASA  programs for funding this research. L.T. acknowledges helpful discussions with J. Karcz.

\def\jgr{J. Geophys. Res.}
\def\grl{Geophys. Res. Lett.}
\def\nat{Nature}
\def\mnras{Mon. Not. R. Astron. Soc.}
\def\icarus{Icarus}
\def\pasp{Publ. Astron. Soc. Pac.}
\def\ssr{Space Sci. Rev.}

\bibliographystyle{agu}

\end{article}

\end{document}